\documentclass[11pt,eqs]{article}
\usepackage{latexsym,amsmath}

\def\cleq{\setcounter{equation}{0}}
\title{Courant bracket twisted both by a 2-form $B$ and by a bi-vector $\theta$
\thanks{Work supported in part by
the Serbian Ministry of Education and Science, under contract No. 171031.}}
\author{ Lj. Davidovi\'c \thanks{ljubica@ipb.ac.rs}, I. Ivani\v sevi\'c \thanks{ivanisevic@ipb.ac.rs} and B. Sazdovi\'c \thanks{sazdovic@ipb.ac.rs}
\\
{\it Institute of Physics, University of Belgrade}\\
{\it Pregrevica 118, 11080 Belgrade, Serbia}}

\begin{document}
\maketitle
\begin{abstract}
We obtain the Courant bracket twisted simultaneously by a 2-form $B$ and a bi-vector $\theta$ by calculating the Poisson bracket algebra of the symmetry generator in the basis obtained acting with the relevant twisting matrix. It is the extension of the Courant bracket that contains well known Schouten-Nijenhuis and Koszul bracket, as well as some new star brackets. We give interpretation to the star brackets as projections on isotropic subspaces. 
\end{abstract}

\section{Introduction}
\cleq
The Courant bracket \cite{courant, courant1} represents the generalization of the Lie bracket on spaces of generalized vectors, understood as the direct sum of the elements of the tangent bundle and the elements of the cotangent bundle. It was obtained in the algebra of generalized currents firstly in \cite{c}. Generalized currents are arbitrary functionals of the fields, parametrized by a pair of vector field and covector field on the target space. Although the Lie bracket satisfies the Jacobi identity, the Courant bracket does not. 

In bosonic string theory, the Courant bracket is governing both local gauge and general coordinate transformations, invariant upon T-duality \cite{doucou, dft}. It is a special case of the more general $C$-bracket \cite{siegel1, siegel}. The $C$-bracket is obtained as the T-dual invariant bracket of the symmetry generator algebra, when the symmetry parameters depend both on the initial and T-dual coordinates. It reduces to the Courant bracket once when parameters depend solely on the coordinates from the initial theory. 

It is possible to obtain the twisted Courant bracket, when the self T-dual generator algebra is considered in the basis obtained from the action of the appropriate $O(D,D)$ transformation \cite{cdual}. The Courant bracket is usually twisted by a 2-form $B$, giving rise to what is known as the twisted Courant bracket \cite{twist}, and by a bi-vector $\theta$, giving rise to the $\theta$-twisted Courant bracket \cite{royt}. In \cite{c, cdual, nick1, nick2}, the former bracket was obtained in the generalized currents algebra, and it was shown to be related to the the latter by self T-duality \cite{crdual}, when the T-dual of the $B$ field is the bi-vector $\theta$. 

The $B$-twisted Courant bracket contains $H$ flux, while the $\theta$-twisted Courant bracket contains non-geometric $Q$ and $R$ fluxes. The fluxes are known to play a crucial role in the compactification of additional dimensions in string theory \cite{granaflux}. Non-geometric fluxes can be used to stabilize moduli. In this paper, we are interested in obtaining the Poisson bracket representation of the twisted Courant brackets that contain all fluxes from the generators algebra. Though it is possible to obtain various twists of the $C$-bracket as well \cite{twistC}, we do not deal with them in this paper.

The realization of all fluxes using the generalized geometry was already considered, see \cite{flux1} for a comprehensive review. In \cite{flux2}, one considers the generalized tetrads originating from the generalized metric of the string Hamiltonian. As the Lie algebra of tetrads originating from the initial metric defines the geometric flux, it is suggested that all the other fluxes can be extracted from the Courant bracket of the generalized tetrads. Different examples of $O(D, D)$ and $O(D) \times O(D)$ transformations of generalized tetrads lead to the Courant bracket algebras with different fluxes as its structure constants.

In \cite{flux3}, one considers the standard Lie algebroid defined with the Lie bracket and the identity map as an anchor on the tangent bundle, as well as the Lie algebroid with the Koszul bracket and the bi-vector $\theta$ as an anchor on the cotangent bundle. The tetrad basis in these Lie algebroids is suitable for defining the geometric $f$ and non-geometric $Q$ fluxes. It was shown that by twisting both of these Lie algebroids by $H$-flux one can construct the Courant algebroid, which gives rise to all of the fluxes in the Courant bracket algebra. Unlike previous approaches where generalized fluxes were defined using the Courant bracket algebra, in a current paper we obtain them in the Poisson bracket algebra of the symmetry generator.

Firstly, we consider the symmetry generator of local gauge and global coordinate transformations, defined as a standard inner product in the generalized tangent bundle of a double gauge parameter and a double canonical variable. The $O(D,D)$ group transforms the double canonical variable into some other basis, in terms of which the symmetry generator can be expressed.
We demonstrate how the Poisson bracket algebra of this generator can be used to obtain twist of the Courant bracket by any such transformation. We give a brief summary of how $e^{\hat{B}}$ and $e^{\hat{\theta}}$ produce respectively the $B$-twisted and $\theta$-twisted Courant bracket in the Poisson bracket algebra of generators \cite{cdual}.

Secondly, we consider the matrix $e^{\breve{B}}$ used for twisting the Courant bracket simultaneously by a 2-form and a bi-vector. The argument $\breve{B}$ is defined simply as a sum of the arguments $\hat{B}$ and $\hat{\theta}$. Unlike $\hat{B}$ or $\hat{\theta}$, the square of $\breve{B}$ is not zero. The full Taylor series gives rise to the hyperbolic functions of the parameter depending on the contraction of the 2-form with the bi-vector $\alpha^{\mu}_{\ \nu} = 2\kappa \theta^{\mu \rho} B_{\rho \nu}$. We represent the symmetry generator in the basis obtained acting with the twisting matrix $e^{\breve{B}}$ on the double canonical variable. This generator is manifestly self T-dual and its algebra closes on the Courant bracket twisted by both $B$ and $\theta$. 

Instead of computing the $B-\theta$ twisted Courant bracket directly, we introduce the change of basis in which we define some auxiliary generators, in order to simplify the calculations. This change of basis is also realized by the action of an element of the $O(D,D)$ group. The structure constants appearing in the Poisson bracket algebra have exactly the same form as the generalized fluxes obtained in other papers \cite{flux1, flux2, flux3}. The expressions for fluxes is given in terms of new auxiliary fields $\mathring{B}$ and $\mathring{\theta}$, both being the function of $\alpha^{\mu}$.  

The algebra of these new auxiliary generators closes on another bracket, that we call $\mathring{C}$-twisted Courant bracket. We obtain its full Poisson bracket representation, and express it in terms of generalized fluxes. We proceed with rewriting it in the coordinate free notation, where many terms are recognized as the well known brackets, such as the Koszul or Schouten-Nijenhuis bracket, but some new brackets, that we call star brackets, also appear. These star brackets as a domain take the direct sum of tangent and cotangent bundle, and as a result give the graph of the bi-vector $\mathring{\theta}$ in the cotangent bundle, i.e. the sub-bundle for which the vector and 1-form components are related as $\xi^\mu = \kappa \mathring{\theta}^{\mu \nu} \lambda_\nu$. We show that they can be defined in terms of the projections on isotropic subspaces acting on different twists of the Courant bracket. 

Lastly, we return to the previous basis and obtain the full expression for the Courant bracket twisted by both $B$ and $\theta$. It has a similar form as $\mathring{C}$-twisted Courant bracket, but in this case the other brackets contained within it are also twisted. The Courant bracket twisted by both $B$ and $\theta$ and the one twisted by $\mathring{C}$ are directly related by a $O(D,D)$ transformation represented with the block diagonal matrix.

\section{The bosonic string essentials}
\cleq
The canonical Hamiltonian for closed bosonic string, moving in the $D$-dimensional space-time with background characterized by the metric field $G_{\mu \nu}$ and the antisymmetric Kalb-Ramond field $B_{\mu \nu}$ is given by \cite{action, regal}
\begin{equation}
{\cal{H_C}} = \frac{1}{2\kappa} \pi_\mu (G^{-1})^{\mu \nu} \pi_\nu + \frac{\kappa}{2} x^{\prime \mu} G^E_{\mu \nu} x^{\prime \nu} - 2 x^{\prime \mu} B_{\mu \rho} (G^{-1})^{\rho \nu} \pi_\nu \, ,
\end{equation}
where $\pi_\mu$ are canonical momenta conjugate to coordinates $x^\mu$, and 
\begin{equation} \label{eq:Gef}
G^E_{\mu \nu} = G_{\mu \nu} - 4 (B G^{-1} B)_{\mu \nu} \, 
\end{equation}
is the effective metric. The Hamiltonian can be rewritten in the matrix notation
\begin{equation} \label{eq:Hcmat}
{\cal H}_{C} = \frac{1}{2\kappa} (X^T)^M H_{MN} X^N \, , 
\end{equation}
where $X^M $ is a double canonical variable given by 
\begin{equation} \label{eq:Xdouble}
X^M = \begin{pmatrix}
\kappa x^{\prime \mu} \\
\pi_\mu \\
\end{pmatrix}\, ,
\end{equation}
and $H_{MN}$ is the so called generalized metric, given by
\begin{equation} \label{eq:genmet}
H_{MN} = 
\begin{pmatrix}
G^E_{\mu \nu} & - 2B_{\mu \rho} (G^{-1})^{\rho \nu} \\
2(G^{-1})^{\mu \rho} B_{\rho \nu} & (G^{-1})^{\mu \nu}
\end{pmatrix} \, ,
\end{equation}
with $M,N \in \{ 0,1\}$. In the context of generalized geometry \cite{gualtieri}, the double canonical variable $X^M$ represents the generalized vector. The generalized vectors are $2D$ structures that combine both vector and 1-form components in a single entity. 

The standard T-duality \cite{tdual, tdual1} laws for background fields have been obtained by Buscher \cite{buscher} 
\begin{equation}\label{eq:tdbf}
^\star G^{\mu\nu} =
(G_{E}^{-1})^{\mu\nu}, \quad
^\star B^{\mu\nu} =
\frac{\kappa}{2}
{\theta}^{\mu\nu} \, ,
\end{equation}
where $(G_{E}^{-1})^{\mu\nu}$ is the inverse of the effective metric (\ref{eq:Gef}), and $\theta^{\mu \nu}$ is the non-commutativity parameter, given by
\begin{equation} \label{eq:thetadef}
{\theta}^{\mu\nu} = - \frac{2}{\kappa}(G^{-1}_E)^{\mu \rho} B_{\rho \sigma} (G^{-1})^{\sigma \nu} \, .
\end{equation}
The T-duality can be realized without changing the phase space, which is called the self T-duality \cite{crdual}. It has the same transformation rules for the background fields like T-duality (\ref{eq:tdbf}), with additionally interchanging the coordinate $\sigma$-derivatives $\kappa x^{\prime \mu}$ with canonical momenta $\pi_\mu$ 
\begin{equation} \label{eq:xpidual}
\kappa x^{\prime \mu} \cong \pi_\mu \, .
\end{equation}
Since momenta and winding numbers correspond to $\sigma$ integral of respectively $\pi_\mu$ and $\kappa x^{\prime \mu}$, we see that the self T-duality, just like the standard T-duality, swaps momenta and winding numbers.

\subsection{Symmetry generator}

We consider the symmetry generator that at the same time governs the general coordinate transformations, parametrized by $\xi^\mu$, and the local gauge transformations, parametrized by $\lambda_\mu$. The generator is given by \cite{dualsim}  
\begin{equation}\label{eq:gltilde}
G(\xi, \lambda) = \int_0^{2\pi} d\sigma {\cal G} (\xi, \lambda)= \int_0^{2\pi} d\sigma\Big[\xi^\mu\pi_\mu+ \lambda_\mu \kappa x^{\prime\mu} \Big] \, .
\end{equation}
It has been shown that the general coordinate transformations and the local gauge transformations are related by self T-duality \cite{dualsim}, meaning that this generator is self T-dual. If one makes the following change of parameters $\lambda_\mu \to \lambda_\mu + \partial_\mu \varphi$, the generator (\ref{eq:gltilde}) does not change 
\begin{equation} \label{eq:reducible}
G (\xi, \lambda + \partial \varphi) = G (\xi, \lambda) + \kappa \int_0^{2\pi}\varphi^\prime d\sigma = G (\xi, \lambda) \, ,
\end{equation} 
since the total derivative integral vanishes for the closed string. Therefore, the symmetry is reducible.

Let us introduce the double gauge parameter $\Lambda^M$, as the generalized vector, given by
\begin{equation} \label{eq:Lxi}
\Lambda^M = \begin{pmatrix}
\xi^\mu \\
\lambda_\mu \\
\end{pmatrix} \, ,
\end{equation}
where $\xi^\mu$ represent the vector components, and $\lambda_\mu$ represent the 1-form components. The space of generalized vectors is endowed with the natural inner product
\begin{equation} \label{eq:skalproizvod}
\langle\Lambda_1,\Lambda_2\rangle = (\Lambda_1^T)^M \eta_{MN} \Lambda_2^N \, \Leftrightarrow \langle (\xi_1, \lambda_1),(\xi_2, \lambda_2)\rangle = i_{\xi_1} \lambda_2 + i_{\xi_2} \lambda_1 = \xi_1^\mu \lambda_{2 \mu} + \xi_2^\mu \lambda_{1 \mu} \, ,
\end{equation}
where $i_{\xi}$ is the interior product along the vector field $\xi$, and $\eta_{MN}$ is $O(D,D)$ metric, given by
\begin{equation} \label{eq:Omegadef}
\eta_{MN} = 
\begin{pmatrix}
0 & 1 \\
1 & 0 
\end{pmatrix} \, .
\end{equation}
Now it is possible to rewrite the generator (\ref{eq:gltilde}) as
\begin{equation} \label{eq:Ggen}
G(\Lambda) = \int d\sigma \langle\Lambda,X\rangle \, .
\end{equation}

In \cite{cdual}, the Poisson bracket algebra of generator (\ref{eq:gltilde}) was obtained in the form
\begin{equation} \label{eq:GGcourant}
\Big\{ G (\Lambda_1), \, G (\Lambda_2) \Big\} = - G \Big([\Lambda_1,\Lambda_2]_{\cal C} \Big) \, ,
\end{equation}
where the standard Poisson bracket relations between coordinates and canonical momenta were assumed 
\begin{equation} \label{eq:PBR}
\{ x^{\mu} (\sigma), \pi_\nu (\bar{\sigma}) \} = \delta^\mu_{\ \nu} \delta(\sigma - \bar{\sigma}) \, .
\end{equation}
The bracket $[\Lambda_1,\Lambda_2]_{\cal C}$ is the Courant bracket \cite{courant}, defined by
\begin{equation}
[\Lambda_1,\Lambda_2]_{\cal C} = \Lambda \Leftrightarrow [(\xi_1, \lambda_1), (\xi_2, \lambda_2)]_{\cal C} = (\xi,\lambda) \, ,
\end{equation}
where
\begin{equation} \label{eq:xicou}
\xi^\mu = \xi_1^\nu \partial_\nu \xi_2^\mu - \xi_2^\nu \partial_\nu \xi_1^\mu \, , \notag
\end{equation}
and
\begin{equation} \label{eq:Lcou}
\lambda_\mu = \xi_1^\nu (\partial_\nu \lambda_{2 \mu} - \partial_\mu \lambda_{2 \nu}) - \xi_2^\nu (\partial_\nu \lambda_{1 \mu} - \partial_\mu \lambda_{1 \nu})+\frac{1}{2} \partial_\mu (\xi_1 \lambda_2- \xi_2 \lambda_1 ) \, .
\end{equation}
It is the generalization of the Lie bracket on spaces of generalized vectors.

\section{$O(D,D)$ group}

Consider the orthogonal transformation ${\cal O}$, i.e. the transformation that preserves the inner product $(\ref{eq:skalproizvod})$ 
\begin{equation} \label{eq:condort}
\langle {\cal O} \Lambda_1, {\cal O} \Lambda_2 \rangle = \langle \Lambda_1, \Lambda_2 \rangle  \Leftrightarrow ({\cal O}\Lambda_1)^T\ \eta\ ({\cal O}\Lambda_2) = \Lambda^T_1\ \eta\ \Lambda_2 \, ,
\end{equation}
which is satisfied for the condition
\begin{equation} \label{eq:condorth}
{\cal O}^T\ \eta\ {\cal O} = \eta \, .
\end{equation}
There is a solution for the above equation in the form ${\cal O} = e^T$, see Sec. 2.1 of \cite{gualtieri}, where
\begin{equation}
T = 
\begin{pmatrix}
A & \theta \\
B & -A^T 
\end{pmatrix} \, ,
\end{equation}
with $\theta : T^\star M \to T M$ and $B: TM \to T^\star M$ being antisymmetric, and $A: TM \to TM $ being the endomorphism. In general case, $B$ and $\theta$ can be independent for ${\cal O}$ to satisfy condition (\ref{eq:condorth}).  

Consider now the action of some element of $O(D,D)$ on the double coordinate $X$ (\ref{eq:Xdouble}) and the double gauge parameter $\Lambda$ (\ref{eq:Lxi})
\begin{equation} \label{eq:hatXL}
\hat{X}^M = {\cal O}^M_{\ N}\ X^N \, ,\ \hat{\Lambda}^M =  {\cal O}^M_{\ N}\ \Lambda^N \, ,
\end{equation}
and note that the relation (\ref{eq:GGcourant}) can be written as
\begin{equation} \label{eq:twdef1}
\int d\sigma \Big\{ \langle \Lambda_1, X \rangle , \langle \Lambda_2, X \rangle \Big\} = - \int d\sigma \langle [\Lambda_1, \Lambda_2]_{\cal C}, X\rangle \, ,
\end{equation}
and using (\ref{eq:condort}) and (\ref{eq:hatXL}) as 
\begin{equation} \label{eq:twdef2}
\int d\sigma \Big\{ \langle \hat{\Lambda}_1, \hat{X} \rangle , \langle \hat{\Lambda}_2,  \hat{X} \rangle \Big\} = - \int d\sigma \langle [\Lambda_1, \Lambda_2]_{\cal C}, X\rangle =-\int d\sigma \langle [\hat{\Lambda}_1, \hat{\Lambda}_2]_{{\cal C}_T}, \hat{X} \rangle    \, ,
\end{equation}
where we expressed the right hand side in terms of some new bracket $[\hat{\Lambda}_1,\hat{\Lambda}_2]_{{\cal C}_T} $. 
Moreover, using (\ref{eq:condort}) and (\ref{eq:hatXL}), the right hand side of (\ref{eq:twdef1}) can be written as
\begin{equation} \label{eq:twdef3}
\langle [\Lambda_1, \Lambda_2]_{\cal C}, X\rangle = \langle [{\cal O}^{-1}\hat{\Lambda}_1, {\cal O}^{-1}\hat{\Lambda}_2]_{\cal C}, {\cal O}^{-1}\hat{X}\rangle =  \langle{\cal O} [{\cal O}^{-1}\hat{\Lambda}_1, {\cal O}^{-1}\hat{\Lambda}_2]_{\cal C}, \hat{X}\rangle \, .
\end{equation}
Using (\ref{eq:twdef2}) and (\ref{eq:twdef3}), one obtains
\begin{equation} \label{eq:twdef}
[\hat{\Lambda}_1,\hat{\Lambda}_2]_{{\cal C}_T} = {\cal O} [{\cal O}^{-1}\hat{\Lambda}_1, {\cal O}^{-1}\hat{\Lambda}_2]_{\cal C} = e^T [e^{-T}\hat{\Lambda}_1, e^{-T}\hat{\Lambda}_2]_{\cal C} \, .
\end{equation}
This is a definition of a $T$-twisted Courant bracket. Throughout this paper, we use the notation where $[,]_{\cal C}$ is the Courant bracket, while when ${\cal C}$ has an additional index, it represents the twist of the Courant bracket by the indexed field, e.g. $[,]_{{\cal C}_B}$ is the Courant bracket twisted by $B$. 

In a special case, when $A= 0$, $\theta = 0$, the bracket (\ref{eq:twdef}) becomes the Courant bracket twisted by a 2-form $B$ \cite{twist}
\begin{equation} \label{eq:CourantB}
[\Lambda_1, \Lambda_2 ]_{{\cal C}_B} = e^{\hat{B}} [e^{-\hat{B}} \Lambda_1, e^{-\hat{B}}\Lambda_2 ]_{\cal C} \, ,
\end{equation}
where $e^{\hat{B}}$ is the twisting matrix, given by
\begin{equation} \label{eq:ebhat}
e^{\hat{B}} = \begin{pmatrix}
\delta^\mu_\nu & 0 \\
2B_{\mu \nu} & \delta^\nu_\mu
\end{pmatrix}\, , \ \ 
\hat{B}^M_{\ N} = 
\begin{pmatrix}
0 & 0 \\
2B_{\mu \nu} & 0 \\ 
\end{pmatrix}\, .
\end{equation}
 This bracket has been obtained in the algebra of generalized currents \cite{nick1, crdual}.

In case of $A=0$, $B=0$, the bracket (\ref{eq:twdef}) becomes the Courant bracket twisted by a bi-vector $\theta$ 
\begin{equation} \label{eq:CourantTheta}
[\Lambda_1, \Lambda_2 ]_{{\cal C}_\theta} = e^{\hat{\theta}} [e^{-\hat{\theta}} \Lambda_1,e^{-\hat{\theta}} \Lambda_2 ]_{\cal C} \, ,
\end{equation}
where $e^{\hat{\theta}}$ is the twisting matrix, given by
\begin{equation} \label{eq:enateta}
e^{\hat{\theta}} = 
\begin{pmatrix}
\delta^\mu_\nu & \kappa \theta^{\mu \nu} \\
0 & \delta^\nu_\mu
\end{pmatrix} \, , \ \ 
\hat{ \theta}^M_{\ N} = 
\begin{pmatrix}
0 & \kappa \theta^{\mu \nu} \\
0 & 0 
\end{pmatrix} \, .
\end{equation}
The $B$-twisted Courant bracket (\ref{eq:CourantB}) and $\theta$-twisted Courant bracket (\ref{eq:CourantTheta}) are related by self T-duality \cite{crdual}. It is easy to demonstrate that both $e^{\hat{B}}$ and $e^{\hat{\theta}}$ satisfy the condition (\ref{eq:condorth}). 

We can now deduce a simple algorithm for finding the Courant bracket twisted by an arbitrary $O(D,D)$ transformation. One rewrites the double symmetry generator $G(\xi,\lambda)$ in the basis obtained by the action of the matrix $e^T$ on the double coordinate (\ref{eq:Xdouble}). Then, the Poisson bracket algebra between these generators gives rise to the appropriate twist of the Courant bracket. In this paper, we apply this algorithm to obtain the Courant bracket twisted by both $B$ and $\theta$. 

\section{Twisting matrix}
\cleq{}

The transformations $e^{\hat{B}}$ and $e^{\hat{\theta}}$ do not commute. That is why we define the transformations that simultaneously twists the Courant bracket by $B$ and $\theta$ as $e^{\breve{B}}$, where 
 \begin{equation} \label{eq:breve}
\breve{B} = \hat{B}+\hat{\theta} = 
\begin{pmatrix}
0 & \kappa \theta^{\mu \nu} \\
2 B_{\mu \nu} & 0
\end{pmatrix} \, .
\end{equation}
The Courant bracket twisted at the same time both by a 2-form $B$ and by a bi-vector $\theta$ is given by
\begin{equation} \label{eq:CTdef}
[\Lambda_1, \Lambda_2 ]_{{\cal C}_{B\theta}} = e^{\breve{B}} [e^{-\breve{B}}\Lambda_1,e^{-\breve{B}} \Lambda_2 ]_{\cal C} \, .
\end{equation}
The full expression for $e^{\breve{B}}$ can be obtained from the well known Taylor series expansion of exponential function
\begin{equation} \label{eq:tayeb}
e^{\breve{B}} = \sum_{n=0}^\infty \frac{\breve{B}^n}{n!} \, .
\end{equation}
The square of the matrix $\breve{B}$ is easily obtained
\begin{equation}
\breve{B}^2 = 
2 
\begin{pmatrix}
\kappa (\theta B)^{\mu}_{\ \nu} & 0 \\
0 & \kappa (B \theta)^{\ \nu}_{\mu}
\end{pmatrix},
\end{equation}
as well as its cube
\begin{equation}
\breve{B}^3 = 
2 
\begin{pmatrix}
0 & \kappa^2 ( \theta B \theta)^{\mu \nu} \\
2\kappa (B\theta B)_{\mu \nu} & 0
\end{pmatrix}.
\end{equation}
The higher degree of $\breve{B}$ are given by
\begin{equation} \label{eq:Bna2n}
\breve{B}^{2n} = 
\begin{pmatrix}
(\alpha^n)^{\mu}_{\ \nu} & 0 \\
0 & ((\alpha^T)^n)^{\ \nu}_{\mu}
\end{pmatrix} \, ,
\end{equation}
for even degrees, and for odd degrees by
\begin{equation} \label{eq:Bna2n1}
\breve{B}^{2n+1} = 
\begin{pmatrix}
0 & \kappa (\alpha^n \theta)^{\mu \nu} \\
2(B \alpha^n )_{\mu \nu} & 0
\end{pmatrix} \, ,
\end{equation}
where we have marked 
\begin{equation} \label{eq:alfadef}
\alpha^{\mu}_{\ \nu} = 2 \kappa \theta^{\mu \rho} B_{\rho \nu}.
\end{equation}

Finally, substituting (\ref{eq:Bna2n}) and (\ref{eq:Bna2n1}) into (\ref{eq:tayeb}), we obtain the twisting matrix
\begin{equation} \label{eq:ebb}
e^{\breve{B}} = 
\begin{pmatrix}
{\cal C}^{\mu}_{\ \nu} & \kappa {\cal S}^\mu_{\ \rho} \theta^{\rho \nu} \\
2 B_{\mu \rho} {\cal S}^{\rho}_{\ \nu} & ( {\cal C}^T)^{\ \nu}_{\mu}
\end{pmatrix} \, ,
\end{equation}
with ${\cal S}^\mu_{\ \nu} = \Big(\frac{\sinh{\sqrt{\alpha}}}{\sqrt{\alpha}} \Big)^\mu_{\ \nu}$ and ${\cal C}^\mu_{\ \nu} = \Big( \cosh{\sqrt{\alpha}} \Big)^{\mu}_{\ \nu}$. 
Its determinant is given by 
\begin{equation}
\det( e^{\breve{B}}) = e^{Tr(\breve{B})} = 1 \, ,
\end{equation}
and the straightforward calculations show that its inverse is given by
\begin{equation} \label{eq:ebmb}
e^{-\breve{B}} =
\begin{pmatrix}
{\cal C}^{\mu}_{\ \nu} & - \kappa{\cal S}^\mu_{\ \rho} \theta^{\rho \nu} \\
-2 B_{\mu \rho} {\cal S}^{\rho}_{\ \nu} & ( {\cal C}^T)^{\ \nu}_{\mu}
\end{pmatrix} \, .
\end{equation}
One easily obtains the relation
\begin{equation}
(e^{\breve{B}})^T\ \eta\ e^{\breve{B}} = \eta \, ,
\end{equation}
therefore the transformation (\ref{eq:ebb}) is indeed an element of $O(D,D)$.

It is worth pointing out characteristics of the matrix $\alpha^\mu_{\ \nu}$. It is easy to show that $\alpha^\mu_{\ \rho} \theta^{\rho \nu} = \theta^{\mu \rho} (\alpha^T)_\rho^{\ \nu}$ and $B_{\mu \rho} \alpha^\rho_{\ \nu} = (\alpha^T)_\mu^{\ \rho} B_{\rho \nu}$, which is further generalized to
\begin{equation} \label{eq:alphaf}
(f(\alpha))^\mu_{\ \rho} \theta^{\rho \nu} = \theta^{\mu \rho} (f(\alpha^T))_\rho^{\ \nu} \, ,\ \ \ B_{\mu \rho} (f(\alpha))^\rho_{\ \nu} = (f(\alpha^T))_\mu^{\ \rho} B_{\rho \nu} \, ,
\end{equation}
for any analytical function $f(\alpha)$.
Moreover, the well known hyperbolic identity $\cosh(x)^2 - \sinh(x)^2 = 1$ can also be expressed in terms of newly defined tensors
\begin{equation} \label{eq:chshid}
({\cal C}^2)^\mu_{\ \nu} - \alpha^\mu_{\ \rho} ({\cal S}^2)^\rho_{\ \nu} = \delta^\mu_\nu \, .
\end{equation}
Lastly, the self T-duality relates the matrix $\alpha$ to its transpose $\alpha \cong \alpha^T$, due to (\ref{eq:tdbf}). Consequently, we write the following self T-duality relations 
\begin{equation} \label{eq:CSdual}
{\cal C} \cong {\cal C}^T \, , \ \ {\cal S} \cong {\cal S}^T \, .
\end{equation} 

\section{Symmetry generator in an appropriate basis}

The direct computation of the bracket (\ref{eq:CTdef}) would be difficult, given the form of the matrix $e^{\breve{B}}$ . Therefore, we use the indirect computation of the bracket, by computing the Poisson bracket algebra of the symmetry generator (\ref{eq:gltilde}), rewritten in the appropriate basis. As elaborated at the end of the Chapter 3, this basis is obtained by the action of the matrix (\ref{eq:ebb}) on the double coordinate (\ref{eq:Xdouble})
\begin{equation} \label{eq:tildeXdef}
\breve{X}^M = (e^{\breve{B}})^M_{\ N} \ X^N =
\begin{pmatrix}
\breve{k}^\mu \\
\breve{\iota}_\mu
\end{pmatrix}
\, ,
\end{equation}
where
\begin{eqnarray} \label{eq:iktilde}
\breve{k}^\mu &=& \kappa {\cal C}^\mu_{\ \nu} x^{\prime \nu} + \kappa ({\cal S} \theta)^{\mu \nu} \pi_\nu \, , \\ \notag
\breve{\iota}_\mu &=& 2  (B{\cal S})_{\mu \nu} x^{\prime \nu} + ( {\cal C}^T)^{\ \nu}_{\mu} \pi_\nu \, ,
\end{eqnarray}
are new currents. Applying (\ref{eq:tdbf}), (\ref{eq:xpidual}) and (\ref{eq:CSdual}) to currents $\breve{k}^\mu$ and $\breve{\iota}_\mu$ we obtain $\breve{\iota}_\mu$ and $\breve{k}^\mu$ respectively, meaning that these currents are directly related by self T-duality. Multiplying the equation (\ref{eq:tildeXdef}) with the matrix (\ref{eq:ebmb}), we obtain the relations inverse to (\ref{eq:iktilde})
\begin{eqnarray} \label{eq:xpiik}
\kappa x^{\prime \mu} &=& {\cal C}^\mu_{\ \nu} \breve{k}^\nu - \kappa ({\cal S} \theta)^{\mu \nu} \breve{\iota}_\nu \, , \\ \notag
\pi_\mu &=& -2 (B{\cal S})_{\mu \nu} \breve{k}^\nu + ({\cal C}^T)^{\ \nu}_{\mu} \breve{\iota}_\nu \, .
\end{eqnarray} 
Applying the transformation (\ref{eq:ebb}) to a double gauge parameter (\ref{eq:Lxi}), we obtain new gauge parameters
\begin{equation} \label{eq:Lxitilde}
\breve{\Lambda}^M = 
\begin{pmatrix}
\breve{\xi}^\mu \\
\breve{\lambda}_\mu
\end{pmatrix} =
(e^{\breve{B}})^M_{\ N}\ \Lambda^N =
\begin{pmatrix}
{\cal C}^\mu_{\ \nu} \xi^\nu + \kappa ({\cal S} \theta)^{\mu \nu} \lambda_\nu \\
2 (B{\cal S})_{\mu \nu} \xi^\nu + ({\cal C}^T)^{\ \nu}_{\mu} \lambda_\nu
\end{pmatrix} \, .
\end{equation}
The symmetry generator (\ref{eq:gltilde}) rewritten in a new basis ${\cal G}({\cal C}\xi+\kappa{\cal S}\theta\lambda , 2(B{\cal S}) \xi + {\cal C}^T \lambda)\equiv{\cal \breve{G}} (\breve{\xi} , \breve{\lambda})$ is given by
\begin{equation} \label{eq:gtilde}
\breve{G}(\breve{\Lambda}) = \int d\sigma \langle\breve{\Lambda},\breve{X} \rangle \Leftrightarrow \breve{G} (\breve{\xi}, \breve{\lambda}) =\int d\sigma\Big[\breve{\xi}^\mu \breve{\iota}_\mu+ \breve{\lambda}_\mu \breve{k}^\mu \Big] \, .
\end{equation}
Substituting (\ref{eq:tildeXdef}) and (\ref{eq:Lxitilde}) into (\ref{eq:gtilde}), the symmetry generator in the initial canonical basis (\ref{eq:gltilde}) is obtained. Due to mutual self T-duality between basis currents (\ref{eq:iktilde}), this generator is invariant upon self T-duality. 

Rewriting the equation (\ref{eq:GGcourant}) in terms of new gauge parameters (\ref{eq:Lxitilde}) in the basis of auxiliary currents (\ref{eq:iktilde}), the Courant bracket twisted by both a 2-form $B_{\mu \nu}$ and by a bi-vector $\theta^{\mu \nu}$ is obtained in the new generator (\ref{eq:gtilde}) algebra
\begin{equation} \label{eq:brbrG}
\Big\{ \breve{G} (\breve{\Lambda}_1), \, \breve{G} (\breve{\Lambda}_2)\Big\} = -\breve{G}\Big( [\breve{\Lambda}_1,\breve{\Lambda}_2]_{{\cal C}_{B\theta}} \Big) \, .
\end{equation}

\subsection{Auxiliary generator}

Let us define a new auxiliary basis, so that both the matrices ${\cal C}$ and ${\cal S}$ are absorbed in some new fields, giving rise to the generator algebra that is much more readable. When the algebra in this basis is obtained, simple change of variables back to the initial ones will provide us with the bracket in need.

Multiplying the second equation of (\ref{eq:iktilde}) with the matrix ${\cal C}^{-1}$, we obtain
\begin{equation} \label{eq:iotaring}
\breve{\iota}_\nu ({\cal C}^{-1})_{\ \mu}^\nu=\pi_\mu + 2 \kappa (B{\cal S} {\cal C}^{-1})_{\mu \nu } x^{\prime \nu}\, ,
\end{equation}
where we have used $(B{\cal S})_{\nu \rho} ({\cal C}^{-1})^\nu_{\ \mu} = - (B {\cal S C}^{-1})_{\rho \mu} = (B {\cal S C}^{-1})_{\mu \rho }$, due to tensor $B{\cal S}$ being antisymmetric, and properties (\ref{eq:alphaf}). 
We will mark the result as a new auxiliary current, given by
\begin{equation} \label{eq:imathring}
\mathring{\iota}_\mu = \pi_\mu + 2 \kappa \mathring{B}_{\mu \nu} x^{\prime \nu} \, ,
\end{equation}
where $\mathring{B}$ is an auxiliary B-field, given by
\begin{equation} \label{eq:Bmathring}
\mathring{B}_{\mu \nu} = B_{\mu \rho} {\cal S}^\rho_{\ \sigma} ({\cal C}^{-1})^\sigma_{\ \nu} \, .
\end{equation}

On the other hand, multiplying the first equation of (\ref{eq:iktilde}) with the matrix ${\cal C}$, we obtain
\begin{equation} \label{eq:kapparing}
{\cal C}^\mu_{\ \nu} \breve{k}^\nu = ({\cal C}^2)^\mu_{\ \nu} \kappa x^{\prime \nu} + \kappa ({\cal C} {\cal S} \theta)^{\mu \nu} \pi_\nu \, .
\end{equation}
Substituting (\ref{eq:chshid}) in the previous equation, and keeping in mind that ${\cal C}$, ${\cal S}$ and $\theta$ commute (\ref{eq:alphaf}), we obtain
\begin{equation} \label{eq:ktrveza}
{\cal C}^\mu_{\ \nu} \breve{k}^\nu = \kappa x^{\prime \mu} + \kappa ({\cal C}{\cal S} \theta)^{\rho \nu}(\pi_\nu + 2\kappa (B {\cal S} {\cal C}^{-1} )_{\nu \sigma} x^{\prime \sigma}) \, .
\end{equation}
Using (\ref{eq:imathring}), the results are marked as a new auxiliary current 
\begin{equation} \label{eq:kmathring}
\mathring{k}^\mu = \kappa x^{\prime \mu} + \kappa \mathring{\theta}^{\mu \nu} \mathring{\iota}_\nu \, , 
\end{equation}
where $\mathring{\theta}$ is given by
\begin{equation} \label{eq:thetamathring}
\mathring{\theta}^{\mu \nu} = {\cal C}^\mu_{\ \rho} {\cal S}^\rho_{\ \sigma} \theta^{\sigma \nu} \, .
\end{equation}
There is no explicit dependence on either ${\cal C}$ nor ${\cal S}$ in redefined auxiliary currents, rather only on canonical variables and new background fields. From (\ref{eq:kmathring}), it is easy to express the coordinate $\sigma$-derivative in the basis of new auxiliary currents
\begin{equation} \label{eq:xikring}
\kappa x^{\prime \mu} = \mathring{k}^\mu - \kappa \mathring{\theta}^{\mu \nu} \mathring{\iota}_\nu \, .
\end{equation}

The first equation of (\ref{eq:iktilde}) could have been multiplied with ${\cal C}$, instead of ${\cal C}^{-1}$, given that the latter would also produce a current that would not explicitly depend on ${\cal C}$. However, the expression for coordinate $\sigma$-derivative $\kappa x^{\prime \mu}$ would explicitly depend on ${\cal C}^2$ in that case, while with our choice of basis it does not (\ref{eq:xikring}).

Substituting (\ref{eq:iotaring}) and (\ref{eq:ktrveza}) in the expression for the generator (\ref{eq:gtilde}), we obtain
\begin{equation}
 \breve{G} (\breve{\xi}, \breve{\lambda}) = \int d\sigma \Big[ \breve{\lambda}_\mu ({\cal C}^{-1})^\mu_{\ \nu} \mathring{k}^\nu + \breve{\xi}^\mu ({\cal C}^T)_\mu^{\ \nu} \mathring{\iota}_\nu \Big] \, ,
\end{equation}
from which it is easily seen that the generator (\ref{eq:gtilde}) is equal to an auxiliary generator
\begin{equation} \label{eq:Gring}
 \mathring{G}(\mathring{\Lambda})= \int d\sigma \langle \mathring{X}, \mathring{\Lambda} \rangle  \Leftrightarrow \mathring{G} (\mathring{\xi}, \mathring{\lambda}) = \int d\sigma \Big[ \mathring{\lambda}_\mu \mathring{k}^\mu + \mathring{\xi}^\mu \mathring{\iota}_\mu \Big] \, ,
\end{equation}
provided that
\begin{equation} \label{eq:vezatildaring}
\mathring{\Lambda}^M = 
\begin{pmatrix}
\mathring{\xi}^\mu \\
\mathring{\lambda}_\mu
\end{pmatrix} \, ,\ \mathring{\lambda}_\mu = \breve{\lambda}_\nu ({\cal C}^{-1})^\nu_{\ \mu},\ \ \ \mathring{\xi}^\mu = {\cal C}^\mu_{\ \nu} \breve{\xi}^\nu \, ,
\end{equation} 
and
\begin{equation} \label{eq:Xring}
\mathring{X}^M = \begin{pmatrix}
\mathring{k}^\mu \\
\mathring{\iota}_\mu 
\end{pmatrix} \, .
\end{equation}
Once that the algebra of (\ref{eq:Gring}) is known, the algebra of generator (\ref{eq:gtilde}) can be easily obtained using (\ref{eq:vezatildaring}).

The change of basis to the one suitable for the auxiliary generator (\ref{eq:Gring}) corresponds to the transformation
\begin{equation} \label{eq:Adef}
A^M_{\ N} = \begin{pmatrix}
({\cal C})^\mu_{\ \nu}   & 0   \\
0 &  (({\cal C}^{-1})^{T})_\mu^{\ \nu}
 \end{pmatrix}  \, , \mathring{\Lambda}^M = A^M_{\ N}\ \breve{\Lambda}^N \, ,\ \mathring{X}^M = A^M_{\ N}\ \breve{X}^N \, ,
\end{equation}
that can be rewritten as
\begin{equation} \label{eq:AeB}
\mathring{X}^M = (A e^{\breve{B}})^M_{\ N}\ X^N \, ,\ \mathring{\Lambda}^M = (A e^{\breve{B}})^M_{\ N}\ \Lambda^N \, ,
\end{equation}
where (\ref{eq:tildeXdef}) and (\ref{eq:Lxitilde}) were used.
It is easy to show that the transformation $A^M_{\ N}$, and consequentially $(A e^{\breve{B}})^M_{\ N}$, is the element of $O(D,D)$ group 
\begin{equation}
A^T \eta\ A = \eta \, , \ (A e^{\breve{B}})^T\ \eta\ (A e^{\breve{B}}) = \eta \, ,
\end{equation}
which means that there is $\mathring{C}$, for which \cite{gualtieri}
\begin{equation} \label{eq:eCCdef}
e^{\mathring{C}} = A e^{\breve{B}} \, .
\end{equation}
The generator (\ref{eq:Gring}) gives rise to algebra that closes on $\mathring{C}$-twisted Courant bracket
\begin{equation}
\Big\{\mathring{G}(\mathring{\Lambda}_1), \mathring{G}(\mathring{\Lambda}_1) \Big\} = - \mathring{G}\Big( [ \mathring{\Lambda}_1,  \mathring{\Lambda}_2]_{{\cal C}_{\mathring{C}}}  \Big) \, ,
\end{equation}
where the $\mathring{C}$-twisted Courant bracket is defined by
\begin{equation} \label{eq:mathCdef}
[ \mathring{\Lambda}_1,  \mathring{\Lambda}_2]_{{\cal C}_{\mathring{C}}} = e^{\mathring{C}} [e^{-\mathring{C}} \mathring{\Lambda}_1, e^{-\mathring{C}} \mathring{\Lambda}_2]_{\cal C} \, .
\end{equation}
In the next chapter, we will obtain this bracket by direct computation of the generators Poisson bracket algebra.

Lastly, let us briefly comment on reducibility conditions for the $\mathring{C}$-twisted Courant bracket. Since we are working with the closed strings, the total derivatives vanishes when integrated out over the worldsheet. Using (\ref{eq:xikring}), we obtain
\begin{equation} \label{eq:ringr}
\int d\sigma \kappa \varphi^\prime = \int d\sigma \kappa x^{\prime \mu} \partial_\mu \varphi = \int d\sigma \Big( \mathring{k}^\mu \partial_\mu \varphi + \kappa \mathring{\iota}_\mu \mathring{\theta}^{\mu \nu} \partial_\nu \varphi \Big) = 0 \, ,
\end{equation}
for any parameter $\lambda$. Hence, the generator (\ref{eq:Gring}) remains invariant under the following change of parameters
\begin{equation}\label{eq:ringred}
\mathring{\xi}^\mu \to \mathring{\xi}^\mu + \kappa \mathring{\theta}^{\mu \nu} \partial_\nu \varphi, \ \ \mathring{\lambda}_\mu \to \mathring{\lambda}_\mu + \partial_\mu \varphi \, .
\end{equation}
These are reducibility conditions (\ref{eq:reducible}) in the basis spanned by $\mathring{k}^\mu$ and $\mathring{\iota}_\mu$.

\section{Courant bracket twisted by $\mathring{C}$ from the generator algebra}
\cleq	
In order to obtain the Poisson bracket algebra for the generator (\ref{eq:Gring}), let us firstly calculate the algebra of basis vectors, using the standard Poisson bracket relations (\ref{eq:PBR}). The auxiliary currents $\mathring{\iota}_\mu$ algebra is 
\begin{equation} \label{eq:iotaiota}
\{ \mathring{\iota}_\mu (\sigma), \mathring{\iota}_\nu (\bar{\sigma})\} = - 2\mathring{B}_{\mu \nu \rho} \mathring{k}^{\rho} \delta(\sigma-\bar{\sigma})- {\cal \mathring{F}}_{\mu \nu}^{\ \rho}\   \mathring{\iota}_\rho \delta(\sigma-\bar{\sigma}) \, ,
\end{equation}
where $\mathring{B}_{\mu \nu \rho}$ is the generalized H-flux, given by
\begin{equation} \label{eq:calB}
\mathring{B}_{\mu \nu \rho} = \partial_\mu \mathring{B}_{\nu \rho} + \partial_\nu \mathring{B}_{\rho \mu}+\partial_\rho \mathring{B}_{\mu \nu} \, ,
\end{equation}
and ${\cal \mathring{F}}_{\mu \nu}^{\rho}$ is the generalized f-flux, given by
\begin{equation} \label{eq:calF}
{\cal \mathring{F}}_{\mu \nu}^{\ \rho} = -2\kappa \mathring{B}_{\mu \nu \sigma} \mathring{\theta}^{\sigma \rho} \, .
\end{equation}
The algebra of currents $\mathring{k}^\mu$ is given by
\begin{eqnarray} \label{eq:kringkring}
\{ \mathring{k}^\mu (\sigma), \mathring{k}^\nu (\bar{\sigma}) \} = - \kappa{\cal \mathring{Q}}_{\rho}^{\ \mu \nu} \mathring{k}^\rho \delta(\sigma-\bar{\sigma}) -\kappa^2{\cal \mathring{R}}^{\mu \nu \rho} \mathring{\iota}_\rho \delta(\sigma-\bar{\sigma}) \, ,
\end{eqnarray}
where
\begin{equation} \label{eq:calQ}
{\cal \mathring{Q}}^{\ \nu \rho}_\mu
 =   \mathring{Q}^{\ \nu \rho}_\mu + 2\kappa \mathring{\theta}^{\nu\sigma} \mathring{\theta}^{\rho \tau} \mathring{B}_{\mu \sigma \tau} \, , \quad  \mathring{Q}^{\ \nu \rho}_\mu = \partial_{\mu} \mathring{\theta}^{\nu \rho}
\end{equation}
and
\begin{equation} \label{eq:calR}
{\cal \mathring{R}}^{\mu \nu \rho} = \mathring{R}^{\mu \nu \rho} +2\kappa \mathring{\theta}^{\mu \lambda} \mathring{\theta}^{\nu \sigma}\mathring{\theta}^{\rho \tau}\mathring{B}_{\lambda \sigma \tau} \, , \quad \mathring{R}^{\mu \nu \rho} = \mathring{\theta}^{\mu \sigma}\partial_\sigma \mathring{\theta}^{\nu \rho} + \mathring{\theta}^{\nu \sigma}\partial_\sigma \mathring{\theta}^{\rho \mu}+ \mathring{\theta}^{\rho \sigma}\partial_\sigma \mathring{\theta}^{\mu \nu} \, .
\end{equation}
The terms in (\ref{eq:kringkring}) containing both $\mathring{\theta}$ and $\mathring{B}$ are the consequence of non-commutativity of auxiliary currents $\mathring{\iota}_\mu$. The remaining algebra of currents $\mathring{k}^\mu$ and $\mathring{\iota}_\mu$ can be as easily obtained
\begin{equation} \label{eq:kringiota}
\{ \mathring{\iota}_\mu (\sigma), \mathring{k}^\nu (\bar{\sigma}) \} = \kappa \delta_\mu^\nu \delta^\prime(\sigma-\bar{\sigma}) + {\cal \mathring{F}}^{\ \nu}_{\mu \rho}\ \mathring{k}^\rho \delta(\sigma-\bar{\sigma}) -\kappa {\cal \mathring{Q}}_{\mu}^{\ \nu \rho} \mathring{\iota}_\rho \delta(\sigma-\bar{\sigma}) \, .
\end{equation}

The basic algebra relations can be summarized in a single algebra relation where the structure constants contain all generalized fluxes 
\begin{equation}
\{\mathring{X}^M , \mathring{X}^N \} = -\mathring{F}^{MN}_{\ \ \ \ P}\ \mathring{X}^P \delta(\sigma-\bar{\sigma}) + \kappa \eta^{MN} \delta^{\prime}(\sigma-\bar{\sigma}) \, ,
\end{equation}
with 
\begin{eqnarray}\label{eq:Fdef}
 F^{M N \rho} =
\begin{pmatrix}
  \kappa^2  \mathring{{\cal R}}^{\mu \nu \rho}  & - \kappa \mathring{{\cal Q}}_\nu^{\ \mu \rho} \\
\kappa \mathring{{\cal Q}}_\mu^{\ \nu \rho}   &  \mathring{{\cal F}}^{\ \rho}_{\mu \nu}  \\
\end{pmatrix}  \,  , \qquad
 F^{M N}{}_\rho  =
\begin{pmatrix}
  \kappa \mathring{{\cal Q}}_\rho^{\ \mu \nu}   &  \mathring{{\cal F}}^{\ \mu}_{\nu \rho }  \\
 -\mathring{{\cal F}}^{\ \nu}_{\mu \rho}   &   2 \mathring{B}_{\mu \nu \rho} \\
\end{pmatrix} \, .
 \end{eqnarray}

The form of the generalized fluxes is the same as the ones already obtained using the tetrad formalism \cite{flux1, flux2, flux3}. In our approach, the generalized fluxes are obtained in the Poisson bracket algebra, only from the fact that the generalized canonical variable $X^M$ is transformed with an element of the $O(D,D)$ group that twists the Courant bracket both by $B$ and $\theta$ at the same time. Consequentially, the fluxes obtained in this paper are functions of some new effective fields, $\mathring{B}_{\mu \nu}$ (\ref{eq:Bmathring}) and $\mathring{\theta}^{\mu \nu}$ (\ref{eq:thetamathring}).

We now proceed to obtain the full bracket. Let us rewrite the generator (\ref{eq:Gring}) algebra
\begin{eqnarray} \label{eq:algdeodeo}
&& \Big\{  \mathring{G} (\mathring{\xi}_1, \mathring{\lambda}_1) (\sigma), \, \mathring{G} (\mathring{\xi}_2, \mathring{\lambda}_2)(\bar{\sigma}) \Big\} = \\ \notag
&& \int d\sigma d\bar{\sigma} \Big[\Big\{\mathring{\xi}_1^\mu(\sigma) \mathring{\iota}_\mu (\sigma), \mathring{\xi}_2^\nu(\bar{\sigma}) \mathring{\iota}_\nu (\bar{\sigma}) \Big\} +\Big\{\mathring{\lambda}_{1\mu}(\sigma) \mathring{k}^\mu (\sigma), \mathring{\lambda}_{2\nu}(\bar{\sigma}) \mathring{k}^\nu (\bar{\sigma}) \Big\} \\ \notag
&&+\Big\{\mathring{\xi}^\mu_1(\sigma) \mathring{\iota}_\mu (\sigma), \mathring{\lambda}_{2\nu}(\bar{\sigma}) \mathring{k}^\nu (\bar{\sigma}) \Big\}+\Big\{\mathring{\lambda}_{1\mu}(\sigma) \mathring{k}^\mu (\sigma), \mathring{\xi}_{2}^\nu(\bar{\sigma}) \mathring{\iota}_\nu (\bar{\sigma}) \Big\} 
\Big] \, .
\end{eqnarray}
The first term of (\ref{eq:algdeodeo}) is obtained, using (\ref{eq:iotaiota})
\begin{eqnarray} \label{eq:ii}
&&\int d\sigma d\bar{\sigma} \Big\{\mathring{\xi}_1^\mu(\sigma) \mathring{\iota}_\mu (\sigma), \mathring{\xi}_2^\nu(\bar{\sigma}) \mathring{\iota}_\nu (\bar{\sigma}) \Big\} = \\ \notag 
&& \int d\sigma \Big[ \mathring{\iota}_\mu \Big( \mathring{\xi}_2^\nu \partial_\nu \mathring{\xi}_1^\mu - \mathring{\xi}_1^\nu \partial_\nu \mathring{\xi}_2^\mu- {\cal \mathring{F}}^{\ \mu}_{\nu \rho}\ \mathring{\xi}_1^\nu \mathring{\xi}_2^\rho \Big) -2 \mathring{B}_{\mu \nu \rho} \mathring{k}^\mu \mathring{\xi}_1^\nu \mathring{\xi}_2^\rho \Big] \, .
\end{eqnarray}
The second term is obtained, using (\ref{eq:kringkring})
\begin{eqnarray} \label{eq:kk}
&&\int d\sigma d\bar{\sigma} \Big\{\mathring{\lambda}_{1\mu}(\sigma) \mathring{k}^\mu (\sigma), \mathring{\lambda}_{2\nu}(\bar{\sigma}) \mathring{k}^\nu (\bar{\sigma}) \Big\} = \\ \notag
&& \int d\sigma \Big[ \mathring{k}^\mu \Big( \kappa\mathring{\theta}^{\nu \rho}(\mathring{\lambda}_{2\nu} \partial_\rho \mathring{\lambda}_{1\mu} - \mathring{\lambda}_{1\nu} \partial_\rho \mathring{\lambda}_{2\mu} )-\kappa {\cal \mathring{Q}}_{\mu}^{\ \nu \rho} \mathring{\lambda}_{1\nu} \mathring{\lambda}_{2\rho} \Big)-\mathring{\iota}_{\mu} \kappa^2 {\cal \mathring{R}}^{\mu \nu \rho} \mathring{\lambda}_{1\nu}\mathring{\lambda}_{2\rho} \Big] \, .
\end{eqnarray}
The remaining terms are antisymmetric with respect to $1 \leftrightarrow 2, \ \sigma \leftrightarrow \bar{\sigma}$ interchange. Therefore, it is sufficient to calculate only the first term in the last line of (\ref{eq:algdeodeo})
\begin{eqnarray} \label{eq:ik}
&&\int d\sigma d\bar{\sigma} \Big\{\mathring{\xi}^\mu_1(\sigma) \mathring{\iota}_\mu (\sigma), \mathring{\lambda}_{2\nu}(\bar{\sigma}) \mathring{k}^\nu (\bar{\sigma}) \Big\} = \\ \notag
&& \int d\sigma \Big[ \mathring{k}^\mu \Big(- \mathring{\xi}^\nu_1 \partial_\nu \mathring{\lambda}_{2\mu}-{\cal \mathring{F}}^{\ \nu}_{\mu \rho}\ \mathring{\xi}^\rho_1 \mathring{\lambda}_{2\nu} \Big)+\mathring{\iota}_\mu \Big(\kappa (\mathring{\lambda}_{2\nu} \mathring{\theta}^{\nu \rho})\partial_\rho \mathring{\xi}_1^\mu -\kappa{\cal \mathring{Q}}_\rho^{\ \nu \mu} \mathring{\xi}^\rho_1 \mathring{\lambda}_{2\nu} \Big) \Big] \\ \notag
&&+ \int d\sigma d\bar{\sigma} \kappa \mathring{\xi}^\nu_1(\sigma)\mathring{\lambda}_{2\nu}(\bar{\sigma}) \partial_\sigma \delta(\sigma-\bar{\sigma}) \, .
\end{eqnarray}
In order to transform the anomalous part, we note that
\begin{equation} \label{eq:deltapola}
\partial_\sigma \delta(\sigma-\bar{\sigma}) = \frac{1}{2} \partial_\sigma \delta(\sigma-\bar{\sigma}) -\frac{1}{2} \partial_{\bar{\sigma}} \delta(\sigma-\bar{\sigma}) \, ,
\end{equation}
and
\begin{equation} \label{eq:fdelta}
f(\bar{\sigma}) \partial_\sigma \delta(\sigma-\bar{\sigma}) = f(\sigma) \partial_\sigma \delta(\sigma-\bar{\sigma})+f^\prime (\sigma) \delta(\sigma-\bar{\sigma}) \, .
\end{equation}
Applying (\ref{eq:deltapola}) and (\ref{eq:fdelta}) to the last row of (\ref{eq:ik}), we obtain
\begin{eqnarray} \label{eq:anomtrans}
&& \int d\sigma d\bar{\sigma} \kappa \mathring{\xi}^\nu_1(\sigma)\mathring{\lambda}_{2\nu}(\bar{\sigma}) \partial_\sigma \delta(\sigma-\bar{\sigma}) = \frac{1}{2}\int d\sigma \kappa x^{\prime \mu} \Big(\mathring{\xi}^\nu_1 \partial_\mu \mathring{\lambda}_{2 \nu} - \partial_\mu \mathring{\xi}^\nu_1 \mathring{\lambda}_{2 \nu} \Big)\\ \notag
&&+\frac{\kappa}{2} \int d\sigma d\bar{\sigma} \Big( \mathring{\xi}^\nu_1(\sigma) \mathring{\lambda}_{2 \nu}(\sigma) \partial_\sigma \delta(\sigma-\bar{\sigma})-\mathring{\xi}^\nu_1(\bar{\sigma}) \mathring{\lambda}_{2 \nu}(\bar{\sigma}) \partial_{\bar{\sigma}} \delta(\sigma-\bar{\sigma}) \Big)= \\ \notag
&& \frac{1}{2}\int d\sigma \Big[\mathring{k}^\mu \Big(\mathring{\xi}_1^\nu \partial_\mu \mathring{\lambda}_{2\nu}- \partial_\mu \mathring{\xi}^\nu_1 \mathring{\lambda}_{2 \nu} \Big)+ \mathring{\iota}_\mu\kappa \mathring{\theta}^{\mu\rho } \Big( \mathring{\xi}_1^\nu \partial_\rho \mathring{\lambda}_{2\nu}- \partial_\rho \mathring{\xi}^\nu_1 \mathring{\lambda}_{2 \nu} \Big) \Big] \, ,
\end{eqnarray}
where (\ref{eq:xikring}) was used, as well as antisymmetry of $\mathring{\theta}$. Substituting (\ref{eq:anomtrans}) to (\ref{eq:ik}), we obtain
\begin{eqnarray} \label{eq:iksredjeno}
&&\int d\sigma d\bar{\sigma} \Big\{\mathring{\xi}^\mu_1 (\sigma) \mathring{\iota}_\mu (\sigma), \mathring{\lambda}_{2\nu}(\bar{\sigma}) \mathring{k}^\nu (\bar{\sigma}) \Big\} = \\ \notag
&& \int d\sigma \Big[ \mathring{k}^\mu \Big(\mathring{\xi}^\nu_1 (\partial_\mu \mathring{\lambda}_{2\nu}-\partial_\nu \mathring{\lambda}_{2\mu})-\frac{1}{2}\partial_\mu (\mathring{\xi}_1 \mathring{\lambda}_2)-{\cal \mathring{F}}^{\ \nu}_{\mu \rho}\ \mathring{\xi}^\rho_1 \mathring{\lambda}_{2\nu} \Big) \\ \notag 
&& +\mathring{\iota}_\mu \Big(\kappa (\mathring{\lambda}_{2\nu} \mathring{\theta}^{\nu \rho}) \partial_\rho \mathring{\xi}_1^\mu+ \kappa \mathring{\theta}^{ \mu \rho} \Big( \mathring{\xi}_1^\nu \partial_\rho \mathring{\lambda}_{2\nu}-\frac{1}{2}\partial_\rho (\mathring{\xi}_1 \mathring{\lambda}_2)\Big) -\kappa{\cal \mathring{Q}}^{\ \nu \mu}_{\rho} \mathring{\xi}^\rho_1 \mathring{\lambda}_{2\nu} \Big) \Big] \, . \notag
\end{eqnarray}

Substituting (\ref{eq:ii}), (\ref{eq:kk}) and (\ref{eq:iksredjeno}) into (\ref{eq:algdeodeo}), we write the full algebra of generator in the form
\begin{equation} \label{GringGring}
\Big\{ \mathring{G} (\mathring{\Lambda}_1), \, \mathring{G} (\mathring{\Lambda}_2) \Big\} = -\mathring{G}(\mathring{\Lambda})  \Leftrightarrow \Big\{ \mathring{G} (\mathring{\xi}_1, \mathring{\lambda}_1), \, \mathring{G} (\mathring{\xi}_2, \mathring{\lambda}_2) \Big\} = -\mathring{G}(\mathring{\xi}, \mathring{\lambda}) \, ,
\end{equation}
where
\begin{eqnarray} \label{eq:ringxi}
\mathring{\xi}^\mu &=&\mathring{\xi}_1^\nu \partial_\nu \mathring{\xi}_2^\mu - \mathring{\xi}_2^\nu \partial_\nu \mathring{\xi}_1^\mu- \kappa \mathring{\theta}^{\mu\rho} \Big(\mathring{\xi}_1^\nu \partial_\rho \mathring{\lambda}_{2\nu} - \mathring{\xi}_2^\nu \partial_\rho \mathring{\lambda}_{1\nu}-\frac{1}{2} \partial_\rho (\mathring{\xi}_1 \mathring{\lambda}_2 - \mathring{\xi}_2 \mathring{\lambda}_1 ) \Big) \\ \notag
&& +\kappa \mathring{\theta}^{\nu \rho} (\mathring{\lambda}_{1\nu} \partial_\rho \mathring{\xi}_2^\mu-\mathring{\lambda}_{2\nu}  \partial_\rho \mathring{\xi}_1^\mu) \\ \notag 
&& +\kappa^2 {\cal \mathring{R}}^{\mu \nu \rho}\mathring{\lambda}_{1\nu}\mathring{\lambda}_{2\rho} + {\cal \mathring{F}}^{\ \mu}_{\rho \sigma}\ \mathring{\xi}_1^\rho \mathring{\xi}_2^\sigma +\kappa{\cal \mathring{Q}}_\rho^{\ \nu \mu}(\mathring{\xi}_1^{\rho} \mathring{\lambda}_{2\nu}-\mathring{\xi}_2^{\rho} \mathring{\lambda}_{1\nu})  \, ,
\end{eqnarray}
and
\begin{eqnarray} \label{eq:ringLambda}
\mathring{\lambda}_\mu &=& \mathring{\xi}_1^\nu (\partial_\nu \mathring{\lambda}_{2\mu}-\partial_\mu \mathring{\lambda}_{2\nu}) - \mathring{\xi}_2^\nu (\partial_\nu\mathring{\lambda}_{1\mu}-\partial_\mu \mathring{\lambda}_{1\nu})+\frac{1}{2} \partial_\mu (\mathring{\xi}_1 \mathring{\lambda}_2- \mathring{\xi}_2 \mathring{\lambda}_1 ) \\ \notag
&&+\kappa\mathring{\theta}^{\nu \rho}(\mathring{\lambda}_{1\nu} \partial_\rho \mathring{\lambda}_{2\mu} - \mathring{\lambda}_{2\nu} \partial_\rho \mathring{\lambda}_{1\mu} ) \\ \notag
&&+2\mathring{B}_{\mu \nu \rho} \mathring{\xi}_1^\nu \mathring{\xi}_2^\rho+ \kappa{\cal \mathring{Q}}^{\ \nu \rho}_\mu \mathring{\lambda}_{1\nu}\mathring{\lambda}_{2\rho} +{\cal \mathring{F}}^{\ \nu}_{\mu \sigma}\ \Big( \mathring{\xi}^\sigma_1 \mathring{\lambda}_{2\nu} - \mathring{\xi}^\sigma_2 \mathring{\lambda}_{1\nu})  \, .
\end{eqnarray}

It is possible to rewrite the previous two equations, if we note the relations between the generalized fluxes
\begin{equation}
{\cal \mathring{R}}^{\mu \nu \rho} = \mathring{R}^{\mu \nu \rho}+ \mathring{\theta}^{\mu \sigma} \mathring{\theta}^{\nu \tau} {\cal \mathring{F}}^{\ \rho}_{\sigma \tau} \, , \quad {\cal \mathring{Q}}_\mu^{\ \nu \rho} =\mathring{Q}_\mu^{\ \nu \rho} + \mathring{\theta}^{\nu \sigma} {\cal \mathring{F}}^{\ \rho}_{\mu \sigma} \,  .
\end{equation}
Now we have
\begin{eqnarray} \label{eq:ringxi2}
\mathring{\xi}^\mu &=&\mathring{\xi}_1^\nu \partial_\nu \mathring{\xi}_2^\mu - \mathring{\xi}_2^\nu \partial_\nu \mathring{\xi}_1^\mu \\ \notag
&&+ \kappa \mathring{\theta}^{\mu\rho} \Big(\mathring{\xi}_1^\nu (\partial_\nu \mathring{\lambda}_{2\rho}-\partial_\rho \mathring{\lambda}_{2\nu}) - \mathring{\xi}_2^\nu(\partial_\nu \mathring{\lambda}_{1\rho}- \partial_\rho \mathring{\lambda}_{1\nu})+\frac{1}{2} \partial_\rho (\mathring{\xi}_1 \mathring{\lambda}_2 - \mathring{\xi}_2 \mathring{\lambda}_1 ) \Big) \\ \notag
&& +\kappa \mathring{\xi}_1^\rho \partial_\rho (\mathring{\lambda}_{2\nu} \mathring{\theta}^{\nu \mu}) -\kappa (\mathring{\lambda}_{2\nu} \mathring{\theta}^{\nu \rho}) \partial_\rho \mathring{\xi}_1^\mu-\kappa \mathring{\xi}_2^\rho \partial_\rho (\mathring{\lambda}_{1\nu} \mathring{\theta}^{\nu \mu})+\kappa(\mathring{\lambda}_{1\nu} \mathring{\theta}^{\nu \rho}) \partial_\rho \mathring{\xi}_2^\mu +\kappa^2 \mathring{R}^{\mu \nu \rho} \mathring{\lambda}_{1\nu}\mathring{\lambda}_{2\rho} \\ \notag
&& +{\cal \mathring{F}}^{\ \mu}_{\rho \sigma}\ \mathring{\xi}_1^\rho \mathring{\xi}_2^\sigma +\kappa \mathring{\theta}^{\mu \sigma}{\cal \mathring{F}}_{\sigma \rho}^{\ \nu}\ (\mathring{\xi}_1^{\rho} \mathring{\lambda}_{2\nu}-\mathring{\xi}_2^{\rho} \mathring{\lambda}_{1\nu}) + \kappa^2 \mathring{\theta}^{\mu \sigma} \mathring{\theta}^{\nu \tau} {\cal \mathring{F}}^{\ \rho}_{\sigma \tau}\ \mathring{\lambda}_{1\nu}\mathring{\lambda}_{2\rho}   \, ,
\end{eqnarray}
and
\begin{eqnarray} \label{eq:ringLambda2}
\mathring{\lambda}_\mu &=& \mathring{\xi}_1^\nu (\partial_\nu \mathring{\lambda}_{2\mu}-\partial_\mu \mathring{\lambda}_{2\nu}) - \mathring{\xi}_2^\nu (\partial_\nu\mathring{\lambda}_{1\mu}-\partial_\mu \mathring{\lambda}_{1\nu})+\frac{1}{2} \partial_\mu (\mathring{\xi}_1 \mathring{\lambda}_2- \mathring{\xi}_2 \mathring{\lambda}_1 ) \\ \notag
&&+\kappa\mathring{\theta}^{\nu \rho}(\mathring{\lambda}_{1\nu} \partial_\rho \mathring{\lambda}_{2\mu} - \mathring{\lambda}_{2\nu} \partial_\rho \mathring{\lambda}_{1\mu} )+\kappa \mathring{Q}_\mu^{\ \nu \rho} \mathring{\lambda}_{1\nu} \mathring{\lambda}_{2\rho} \\ \notag
&&+2\mathring{B}_{\mu \nu \rho} \mathring{\xi}_1^\nu \mathring{\xi}_2^\rho+ {\cal \mathring{F}}^{\ \nu}_{\mu \sigma}\ \Big( \mathring{\xi}^\sigma_1 \mathring{\lambda}_{2\nu} - \mathring{\xi}^\sigma_2 \mathring{\lambda}_{1\nu})+ \kappa \mathring{\theta}^{\nu \sigma} {\cal \mathring{F}}_{ \mu \sigma}^{\ \rho}\ \mathring{\lambda}_{1\nu}\mathring{\lambda}_{2\rho}    \, ,
\end{eqnarray}
where the partial integration was used in the equation (\ref{eq:ringxi2}). 

The relation (\ref{GringGring}) defines the $\mathring{C}$-twisted Courant bracket 
\begin{equation} \label{eq:bracket4}
[\mathring{\Lambda}_1, \mathring{\Lambda}_2]_{{\cal C}_{\mathring{C}}} = \mathring{\Lambda} \Leftrightarrow [(\mathring{\xi}_1,\mathring{\lambda}_1), (\mathring{\xi}_2,\mathring{\lambda}_2) ]_{{\cal C}_{\mathring{C}}} = (\mathring{\xi},\mathring{\lambda}) \, ,
\end{equation}
that gives the same bracket as (\ref{eq:mathCdef}). Both (\ref{eq:ringxi}) - (\ref{eq:ringLambda}) and (\ref{eq:ringxi2}) - (\ref{eq:ringLambda2}) are the products of $\mathring{C}$-twisted Courant bracket. The former shows explicitly how the gauge parameters depend on the generalized fluxes. In the latter, similarities between the expressions for two parameters is easier to see. 

\subsection{Special cases and relations to other brackets}

Even though the non-commutativity parameter $\theta$ and the Kalb Ramond field $B$ are not mutually independent, while obtaining the bracket (\ref{eq:bracket4}) the relation between these fields (\ref{eq:thetadef}) was not used. Therefore, the results stand even if a bi-vector and a 2-form used for twisting are mutually independent. This will turn out to be convenient to analyze the origin of terms appearing in the Courant bracket twisted by $\mathring{C}$.

Primarily, consider the case of zero bi-vector $\theta^{\mu \nu} = 0$ with the 2-form $B_{\mu \nu}$ arbitrary. Consequently, the parameter $\alpha$ (\ref{eq:alfadef}) is zero, while the hyperbolic functions ${\cal C}$ and ${\cal S}$ are identity matrices. Therefore, the auxiliary fields (\ref{eq:Bmathring}) and (\ref{eq:thetamathring}) simplify in a following way
\begin{equation}
\mathring{B}_{\mu \nu} \to B_{\mu \nu} \, \ \ \ \mathring{\theta}^{\mu \nu}\to 0 \, ,
\end{equation}
and the twisting matrix $e^{\breve{B}}$ (\ref{eq:ebb}) becomes the matrix $e^{\hat{B}}$ (\ref{eq:ebhat}). The expressions (\ref{eq:ringxi}) and (\ref{eq:ringLambda}) respectively reduce to 
\begin{equation} \label{eq:XIB}
\mathring{\xi}^\mu = \mathring{\xi}_1^\nu \partial_\nu \mathring{\xi}_2^\mu - \mathring{\xi}_2^\nu \partial_\nu \mathring{\xi}_1^\mu \, ,
\end{equation} and
\begin{equation} \label{eq:LB}
\mathring{\lambda}_\mu = \mathring{\xi}_1^\nu (\partial_\nu \mathring{\lambda}_{2 \mu} - \partial_\mu \mathring{\lambda}_{2 \nu}) - \mathring{\xi}_2^\nu (\partial_\nu \mathring{\lambda}_{1 \mu} - \partial_\mu \mathring{\lambda}_{1 \nu}) +\frac{1}{2} \partial_\mu (\mathring{\xi}_1 \mathring{\lambda}_2- \mathring{\xi}_2 \mathring{\lambda}_1 )+ 2 B_{\mu \nu \rho} \mathring{\xi}^\nu_1 \mathring{\xi}^\rho_2 \, ,
\end{equation}
where $B_{\mu \nu \rho}$ is ithe Kalb-Ramond field strength, given by 
\begin{equation} \label{eq:bmnr}
B_{\mu \nu \rho} = \partial_\mu B_{\nu \rho} + \partial_\nu B_{\rho \mu} + \partial_\rho B_{\mu \nu} \, .
\end{equation} 
The equations (\ref{eq:XIB}) and (\ref{eq:LB}) define exactly the $B$-twisted Courant bracket (\ref{eq:CourantB}) \cite{twist}. 

Secondarily, consider the case of zero 2-form $B_{\mu \nu} = 0$ and the bi-vector $\theta^{\mu \nu}$ arbitrary. Similarly, $\alpha=0$ and ${\cal C}$ and ${\cal S}$ are identity matrices. The auxiliary fields $\mathring{B}_{\mu \nu}$ and $\mathring{\theta}^{\mu \nu}$ are given by
\begin{equation}
\mathring{B}_{\mu \nu} \to 0 \, \ \ \ \mathring{\theta}^{\mu \nu} \to \theta^{\mu \nu} \, .
\end{equation}
The twisting matrix $e^{\breve{B}}$ becomes the matrix of $\theta$-transformations $e^{\hat{\theta}}$ (\ref{eq:enateta}). The gauge parameters (\ref{eq:ringxi}) and (\ref{eq:ringLambda}) are respectively given by
\begin{align} \label{eq:XIR}
\mathring{\xi}^\mu =&\ \mathring{\xi}_1^\nu \partial_\nu \mathring{\xi}_2^\mu - \mathring{\xi}_2^\nu \partial_\nu\mathring{\xi}_1^\mu + \\ \notag
& +\kappa \theta^{\mu \rho}\Big( \mathring{\xi}_1^\nu (\partial_\nu \mathring{\lambda}_{2 \rho}-\partial_\rho \mathring{\lambda}_{2 \nu}) - \mathring{\xi}_2^\nu ( \partial_\nu \mathring{\lambda}_{1 \rho}-\partial_\rho \mathring{\lambda}_{1 \nu}) +\frac{1}{2} \partial_\rho (\mathring{\xi}_1 \mathring{\lambda}_{2} - \mathring{\xi}_2 \mathring{\lambda}_1) \Big) \\ \notag
& + \kappa \mathring{\xi}_1^\nu \partial_\nu (\mathring{\lambda}_{2 \rho} \theta^{\rho \mu})-\kappa \mathring{\xi}_2^\nu \partial_\nu (\mathring{\lambda}_{1 \rho} \theta^{\rho \mu})+\kappa (\mathring{\lambda}_{1 \nu} \theta^{\nu \rho}) \partial_\rho \mathring{\xi}_2^\mu -\kappa (\mathring{\lambda}_{2 \nu}\theta^{\nu \rho}) \partial_\rho \mathring{\xi}_1^\mu \\ \notag
&+\kappa^2 R^{\mu \nu \rho} \mathring{\lambda}_{1 \nu}\mathring{\lambda}_{2 \rho} \, ,
\end{align}
and
\begin{align} \label{eq:LR}
\mathring{\lambda}_\mu = &\ \mathring{\xi}_1^\nu (\partial_\nu \mathring{\lambda}_{2 \mu} - \partial_\mu \mathring{\lambda}_{2 \nu}) - \mathring{\xi}_2^\nu (\partial_\nu \mathring{\lambda}_{1 \mu} - \partial_\mu \mathring{\lambda}_{1 \nu}) +\frac{1}{2}\partial_\mu(\mathring{\xi}_1 \mathring{\lambda}_2 - \mathring{\xi}_2 \mathring{\lambda}_1) \\ \notag
& + \kappa \theta^{\nu \rho} (\mathring{\lambda}_{1 \nu}\partial_\rho \mathring{\lambda}_{2 \mu}-\mathring{\lambda}_{2 \nu} \partial_\rho \mathring{\lambda}_{1 \mu})+ \kappa \mathring{\lambda}_{1 \rho} \mathring{\lambda}_{2 \nu} Q_\mu^{\ \rho \nu} \, ,
\end{align}
where by $Q_\mu^{\ \nu \rho}$ and $R^{\mu \nu \rho}$ we have marked the non-geometric fluxes, given by
\begin{equation} \label{eq:nongeomflux}
Q_\mu^{\ \nu \rho} = \partial_\mu \theta^{\nu \rho},\ \ R^{\mu \nu \rho} = \theta^{\mu \sigma} \partial_\sigma \theta^{\nu \rho} + \theta^{\nu \sigma} \partial_\sigma \theta^{\rho \mu} +\theta^{\rho\sigma} \partial_\sigma \theta^{\mu \nu} \, .
\end{equation}
The bracket defined by these relations is $\theta$-twisted Courant bracket (\ref{eq:CourantTheta}) \cite{cdual} and it features the non-geometric fluxes only. 

Let us comment on terms in the obtained expressions for gauge parameters (\ref{eq:ringxi2}) and (\ref{eq:ringLambda2}). The first line of (\ref{eq:ringxi2}) appears in the Courant bracket and in all brackets that can be obtained from its twisting by either a 2-form or a bi-vector. The next two lines correspond to the terms appearing in the $\theta$-twisted Courant bracket (\ref{eq:XIR}). The other terms do not appear in either $B$- or $\theta$-twisted Courant bracket.

Similarly, the first line of (\ref{eq:ringLambda}) appears in the Courant bracket (\ref{eq:Lcou}) and in all other brackets obtained from its twisting, while the terms in the second line appear exclusively in the $\theta$ twisted Courant bracket (\ref{eq:LB}). The first term in the last line appear in the $B$-twisted Courant bracket (\ref{eq:LR}), while the rest are some new terms. We see that all the terms that do not appear in neither of two brackets are the terms containing ${\cal \mathring{F}}$ flux.

\subsection{Coordinate free notation}

In order to obtain the formulation of the $\mathring{C}$-twisted Courant bracket in the coordinate free notation, independent of the local coordinate system that is used on the manifold, let us firstly provide definitions for a couple of well know brackets and derivatives. 

The Lie derivative along the vector field $\xi$ is given by
\begin{equation} \label{eq:lieder}
{\cal L}_{\mathring{\xi}} =i_{\mathring{\xi}} d + d i_{\mathring{\xi}} \, ,
\end{equation}
with $i_{\mathring{\xi}}$ being the interior product along the vector field $\mathring{\xi}$ and $d$ being the exterior derivative. Using the Lie derivative one easily defines the Lie bracket
\begin{eqnarray} \label{eq:liebr}
[\mathring{\xi}_1, \mathring{\xi}_2]_L = {\cal L}_{\mathring{\xi}_1} {\mathring{\xi}_2} - {\cal L}_{\mathring{\xi}_2} {\mathring{\xi}_1} \, .
\end{eqnarray}
The generalization of the Lie bracket on a space of 1-forms is a well known Koszul bracket \cite{koszul}
\begin{equation} \label{eq:koszul}
[\mathring{\lambda}_1, \mathring{\lambda}_2]_\theta = {\cal{L}}_{\mathring{\theta} \mathring{\lambda}_1 } \mathring{\lambda}_2 - {\cal{L}}_{ \mathring{\theta}\mathring{\lambda}_2} \mathring{\lambda}_1 + d(\mathring{\theta}(\mathring{\lambda}_1, \mathring{\lambda}_2)) \, .
\end{equation} 

The expressions (\ref{eq:ringxi}) and (\ref{eq:ringLambda}) in the coordinate free notation are given by
\begin{eqnarray} \label{eq:ringRxi}
\mathring{\xi} &=& [\mathring{\xi}_1,\mathring{\xi}_2]_L - [\mathring{\xi}_2,\mathring{\lambda}_1 \kappa \mathring{\theta}]_L + [\mathring{\xi}_1,\mathring{\lambda}_2 \kappa \mathring{\theta}]_L \\ \notag
&&- \Big({\cal L}_{\mathring{\xi}_1}\mathring{\lambda}_2 - {\cal L}_{\mathring{\xi}_2}\mathring{\lambda}_1 - \frac{1}{2}d(i_{\mathring{\xi}_1}\mathring{\lambda}_2 - i_{\mathring{\xi}_2}\mathring{\lambda}_1)\Big)\kappa \mathring{\theta}\\ \notag
&&+ {\cal \mathring{F}}(\mathring{\xi}_1,\mathring{\xi}_2, .) - \kappa \mathring{\theta}{\cal \mathring{F}}(\mathring{\lambda}_1, ., \mathring{\xi}_2)  +\kappa \mathring{\theta}{\cal \mathring{F}} (\mathring{\lambda}_2, ., \mathring{\xi}_1) + {\cal \mathring{R}} (\mathring{\lambda}_1,\mathring{\lambda}_2,.) \, ,
\end{eqnarray}
and
\begin{eqnarray} \label{eq:ringRLambda}
\mathring{\lambda} &=& {\cal L}_{\mathring{\xi}_1}\mathring{\lambda}_2 - {\cal L}_{\mathring{\xi}_2}\mathring{\lambda}_1 - \frac{1}{2}d(i_{\mathring{\xi}_1}\mathring{\lambda}_2 - i_{\mathring{\xi}_2}\mathring{\lambda}_1) - [\mathring{\lambda}_1,\mathring{\lambda}_2]_{\kappa \mathring{\theta}} \\ \notag
&&+ \mathring{H}(\mathring{\xi}_1,\mathring{\xi}_2,.)-{\cal \mathring{F}} (\mathring{\lambda}_1, ., \mathring{\xi}_2) +{\cal \mathring{F}} (\mathring{\lambda}_2, ., \mathring{\xi}_1)+\kappa\mathring{\theta} {\cal \mathring{F}} (\mathring{\lambda}_1, \mathring{\lambda}_2, .) \, ,
\end{eqnarray} 
where 
\begin{equation} \label{eq:Bnc}
\mathring{H} = 2 d \mathring{B} \, .
\end{equation}
We have marked the geometric $H$ flux as $\mathring{H}$, so that it is distinguished from the 2-form $\mathring{B}$. In the local basis, the full term containing $H$-flux is given by
\begin{equation}
\left. \mathring{H}(\mathring{\xi}_1,\mathring{\xi}_2, .) \right|_\mu = 2\mathring{B}_{\mu \nu \rho }\  \mathring{\xi}_1^{\nu} \mathring{\xi}_2^{\rho}  \, .
\end{equation}
Similarly are defined the terms containing ${\cal \mathring{F}}$ flux
\begin{equation}
\left. {\cal \mathring{F}}(\mathring{\xi}_1,\mathring{\xi}_2, .) \right|^\mu = {\cal \mathring{F}}^{\ \mu}_{\nu \rho}\  \mathring{\xi}_1^{\nu} \mathring{\xi}_2^{\rho} \, ,
\end{equation}
and the non-geometric ${\cal \mathring{R}}$ flux
\begin{equation}
\left. {\cal \mathring{R}} (\mathring{\lambda}_1, \mathring{\lambda}_2,.) \right|^{\mu} =  {\cal \mathring{R}}^{\mu \nu \rho} \mathring{\lambda}_{1\nu} \mathring{\lambda}_{2\rho} \, ,
\end{equation}
as well as 
\begin{equation}
\left. \mathring{\theta} {\cal \mathring{F}} (\mathring{\lambda}_1, . ,  \mathring{\xi}_2) \right|^{\mu} =  \mathring{\theta}^{\nu \sigma}{\cal \mathring{F}}_{\sigma \rho}^{\ \mu}\ \mathring{\lambda}_{1\nu} \mathring{\xi}_{2}^{\rho} \, .
\end{equation}

It is possible to rewrite the coordinate free notation in terms of the $\mathring{H}$-flux and $\mathring{\theta}$ bi-vector only.  The geometric ${\cal \mathring{F}}$ flux is just the contraction of the $\mathring{H}$-flux with a bi-vector
\begin{equation} \label{eq:Fnc}
{\cal \mathring{F}} = \kappa \mathring{\theta}\ \mathring{H} \, .
\end{equation}
The non-geometric ${\cal \mathring{R}}$ flux can be rewritten as
\begin{equation} \label{eq:Rnc}
{\cal \mathring{R}} = \frac{1}{2} [\mathring{\theta}, \mathring{\theta}]_S+ \wedge^3 (\kappa \mathring{\theta}) \mathring{H} \, ,
\end{equation}
 where $\wedge$ is the wedge product, and by $[\mathring{\theta},\mathring{\theta}]_S$ we have marked the Schouten-Nijenhuis bracket \cite{SNB}, given by
\begin{equation} \label{eq:SNb}
\left. [\mathring{\theta}, \mathring{\theta}]_S \right| ^{\mu \nu \rho} = \epsilon^{\mu \nu \rho}_{\alpha \beta \gamma} \mathring{\theta}^{\sigma \alpha} \partial_\sigma \mathring{\theta}^{\beta \gamma} = 2 \mathring{R}^{\mu \nu \rho} \, ,
\end{equation}
where 
\begin{equation}
\epsilon^{\mu \nu \rho}_{\alpha \beta \gamma} = 
\begin{vmatrix}
\delta^\mu_\alpha & \delta^\nu_\beta & \delta^\rho_\gamma \\ 
\delta^\nu_\alpha & \delta^\rho_\beta & \delta^\mu_\gamma \\
\delta^\rho_\alpha & \delta^\mu_\beta & \delta^\nu_\gamma
\end{vmatrix}\, .
\end{equation}
Expressing both ${\cal \mathring{F}}$ and ${\cal \mathring{R}}$ fluxes in terms of the bi-vector $\mathring{\theta}$ and 3-form $\mathring{H}$, we obtain
\begin{eqnarray} \label{eq:ringRxi2}
\mathring{\xi} &=& [\mathring{\xi}_1,\mathring{\xi}_2]_L - [\mathring{\xi}_2,\mathring{\lambda}_1 \kappa \mathring{\theta}]_L + [\mathring{\xi}_1,\mathring{\lambda}_2 \kappa \mathring{\theta}]_L \\ \notag
&&- \Big({\cal L}_{\mathring{\xi}_1}\mathring{\lambda}_2 - {\cal L}_{\mathring{\xi}_2}\mathring{\lambda}_1 - \frac{1}{2}d(i_{\mathring{\xi}_1}\mathring{\lambda}_2 - i_{\mathring{\xi}_2}\mathring{\lambda}_1)\Big)\kappa \mathring{\theta}+ \frac{\kappa^2}{2} [\mathring{\theta},\mathring{\theta}]_S (\mathring{\lambda}_1,\mathring{\lambda}_2,.) \\ \notag
&& +\kappa \mathring{\theta} \mathring{H} (., \mathring{\xi}_1,\mathring{\xi}_2)  - \wedge^2\kappa\mathring{\theta}\mathring{H}(\mathring{\lambda}_1, ., \mathring{\xi}_2)+\wedge^2 \kappa\mathring{\theta} \mathring{H} (\mathring{\lambda}_2, ., \mathring{\xi}_1)+\wedge^3\kappa \mathring{\theta} \mathring{H} (\mathring{\lambda}_1,\mathring{\lambda}_2,.) \, ,
\end{eqnarray}
and
\begin{eqnarray} \label{eq:ringRLambda2}
\mathring{\lambda} &=& {\cal L}_{\mathring{\xi}_1}\mathring{\lambda}_2 - {\cal L}_{\mathring{\xi}_2}\mathring{\lambda}_1 - \frac{1}{2}d(i_{\mathring{\xi}_1}\mathring{\lambda}_2 - i_{\mathring{\xi}_2}\mathring{\lambda}_1) - [\mathring{\lambda}_1,\mathring{\lambda}_2]_{\kappa \mathring{\theta}} \\ \notag
&&+ \mathring{H}(\mathring{\xi}_1,\mathring{\xi}_2,.) -\kappa\mathring{\theta} \mathring{H}(\mathring{\lambda}_1, ., \mathring{\xi}_2)+\kappa\mathring{\theta}\mathring{H}(\mathring{\lambda}_1, . ,\mathring{\xi}_2)+\wedge^2 \kappa\mathring{\theta} \mathring{H}(\mathring{\lambda}_1, \mathring{\lambda}_2,.) \, .
\end{eqnarray} 

The term $\kappa \mathring{\theta} \mathring{H}(. , \mathring{\xi}_1, \mathring{\xi}_2)$ is the wedge product of a bi-vector with a 3-form, contracted with two vectors, given by
\begin{equation}
\Big( \kappa \mathring{\theta} \mathring{H}(. , \mathring{\xi}_1, \mathring{\xi}_2) \Big)^\mu = 2 \kappa \mathring{\theta}^{\mu \nu}\mathring{B}_{\nu \rho \sigma} \mathring{\xi}_{1}^{\rho} \mathring{\xi}_2^\sigma \, ,
\end{equation}
and $\kappa \mathring{\theta}\mathring{H}(\mathring{\lambda}_1,. ,\mathring{\xi}_2) $ is similarly defined, with the 1-form contracted instead of one vector field
\begin{equation}
\Big(\kappa \mathring{\theta}\mathring{H}(\mathring{\lambda}_1,. ,\mathring{\xi}_2) \Big)_\mu = 2 \kappa \mathring{\theta}^{\nu \rho}\mathring{B}_{\rho \mu \sigma } \mathring{\lambda}_{1\nu} \mathring{\xi}_2^\sigma \, .
\end{equation}
The terms like $\wedge^2 \kappa\mathring{\theta} \mathring{H}(\mathring{\lambda}_1, ., \mathring{\xi}_2)$ are the wedge product of two bi-vectors with a 3-form, contracted with the 1-form $\mathring{\lambda}_1$ and the vector $\mathring{\xi}_2$
\begin{equation}
\Big( \wedge^2 \kappa\mathring{\theta}\mathring{H} (\mathring{\lambda}_1, ., \mathring{\xi}_2) \Big)^\mu = 2 \kappa^2 \mathring{\theta}^{\nu \sigma} \mathring{\theta}^{\mu \rho} \ \mathring{B}_{\sigma \rho \tau} \mathring{\lambda}_{1\nu} \mathring{\xi}_2^\tau \, ,
\end{equation}
and similarly when contraction is done with two forms
\begin{equation}
\Big( \wedge^2 \kappa\mathring{\theta}\mathring{H}(\mathring{\lambda}_1, \mathring{\lambda}_2 , .) \Big)_\mu = 2\kappa^2 \mathring{\theta}^{\tau \rho } \mathring{\theta}^{\nu \sigma } \mathring{B}_{\rho \sigma \mu}\mathring{\lambda}_{1\tau} \mathring{\lambda}_{2\nu} \, .
\end{equation}
Lastly, the term $\wedge^3\kappa \mathring{\theta}\mathring{H}(\mathring{\lambda}_1,\mathring{\lambda}_2,.) $ is obtained by taking a wedge product of three bi-vectors with a 3-form and than contracting it with two 1-forms. It is given by
\begin{equation}
\Big(\wedge^3\kappa \mathring{\theta}\mathring{H}(\mathring{\lambda}_1,\mathring{\lambda}_2,.) \Big)^\mu = 2\kappa^3 \mathring{\theta}^{\nu \sigma }\mathring{\theta}^{ \rho \tau}\mathring{\theta}^{\mu \lambda} \mathring{B}_{\sigma \tau \lambda }\mathring{\lambda}_{1\nu}\mathring{\lambda}_{2\rho} \, ,
\end{equation}

\section{Star brackets}
\cleq
The expressions for gauge parameters (\ref{eq:ringRxi}) and (\ref{eq:ringRLambda}) produce some well known bracket, such as Lie bracket and Koszul bracket. The remaining terms can be combined so that they are expressed by some new brackets, acting on pairs of generalized vectors. It turns out that these brackets produce a generalized vector, where the vector part $\mathring{\xi}^\mu$ and the 1-form part $\mathring{\lambda}_\mu$ are related by $\mathring{\xi}^\mu = \kappa \mathring{\theta}^{\mu \nu} \mathring{\lambda}_\nu$, effectively resulting in the graphs in the generalized cotangent bundle $T^\star M$ of the bi-vector $\mathring{\theta}$, i.e. $\xi = \kappa \theta(.,\lambda)$. The star brackets can be interpreted in terms of projections on isotropic subspaces. 

\subsection{$\theta$-star bracket}

Let us firstly consider the second line of (\ref{eq:ringxi2}) and the first line of (\ref{eq:ringLambda2}). When combined, they define a bracket acting on a pair of generalized vectors
\begin{equation} \label{eq:star2}
[\mathring{\Lambda}_1, \mathring{\Lambda}_2]^\star_{\mathring{\theta}} = \mathring{\Lambda}^\star \Leftrightarrow [ (\mathring{\xi}_1, \mathring{\lambda}_1),(\mathring{\xi}_2, \mathring{\lambda}_2)]^\star_{\mathring{\theta}} = (\mathring{\xi}_\star, \mathring{\lambda}^\star) \, ,
\end{equation}
where
\begin{equation} \label{eq:xistar2}
\mathring{\xi}_\star^\mu = \kappa \mathring{\theta}^{\mu\rho} \Big(\mathring{\xi}_1^\nu (\partial_\nu \mathring{\lambda}_{2\rho}-\partial_\rho \mathring{\lambda}_{2\nu}) - \mathring{\xi}_2^\nu(\partial_\nu \mathring{\lambda}_{1\rho}- \partial_\rho \mathring{\lambda}_{1\nu})+ \frac{1}{2} \partial_\rho (\mathring{\xi}_1 \mathring{\lambda}_2- \mathring{\xi}_2 \mathring{\lambda}_1 )\Big)  \, ,
\end{equation}
and
\begin{equation} \label{eq:Lstar2}
\mathring{\lambda}^\star_\mu = \mathring{\xi}_1^\nu (\partial_\nu \mathring{\lambda}_{2\mu}-\partial_\mu \mathring{\lambda}_{2\nu}) - \mathring{\xi}_2^\nu (\partial_\nu\mathring{\lambda}_{1\mu}-\partial_\mu \mathring{\lambda}_{1\nu})+\frac{1}{2} \partial_\mu (\mathring{\xi}_1 \mathring{\lambda}_2- \mathring{\xi}_2 \mathring{\lambda}_1 ) \, ,
\end{equation}
from which one easily reads the relation
\begin{equation} \label{eq:star2veza}
\mathring{\xi}_\star^\mu= \kappa \mathring{\theta}^{\mu \rho} \mathring{\lambda}_\rho^\star \, .
\end{equation}
In a coordinate free notation, this bracket can be written as
\begin{equation}
[ \mathring{\Lambda}_1,\mathring{\Lambda}_2]^\star_{\mathring{\theta}}  = [ (\mathring{\xi}_1, \mathring{\lambda}_1),(\mathring{\xi}_2, \mathring{\lambda}_2)]^\star_{\mathring{\theta}} = \Big(\kappa \mathring{\theta} \Big(.,{\cal L}_{\mathring{\xi}_1}\mathring{\lambda}_2 - {\cal L}_{\mathring{\xi}_2}\mathring{\lambda}_1\Big), {\cal L}_{\mathring{\xi}_1}\mathring{\lambda}_2 - {\cal L}_{\mathring{\xi}_2}\mathring{\lambda}_1 \Big) \, .
\end{equation}

\subsection{$B\theta$-star bracket}

The remaining terms contain geometric $\mathring{H}$ and ${\cal \mathring{F}}$ fluxes. Note that they are the only terms that depend on the new effective Kalb-Ramond field $\mathring{B}$. Firstly, we mark the last line of (\ref{eq:ringLambda2}) as
\begin{equation}  \label{eq:Lstar1}
\mathring{\lambda}^*_{\mu} = 2\mathring{B}_{\mu \nu \rho} \mathring{\xi}_1^\nu \mathring{\xi}_2^\rho+ {\cal \mathring{F}}^{\ \nu}_{\mu \sigma}\ \Big( \mathring{\xi}^\sigma_1 \mathring{\lambda}_{2\nu} - \mathring{\xi}^\sigma_2 \mathring{\lambda}_{1\nu})+ \kappa \mathring{\theta}^{\nu \sigma} {\cal \mathring{F}}_{ \mu \sigma}^{\ \rho}\ \mathring{\lambda}_{1\nu}\mathring{\lambda}_{2\rho} \, .
\end{equation}
Secondly, using the definition of ${\cal \mathring{F}}$ (\ref{eq:calF}) and the fact that $\mathring{\theta}$ is antisymmetric, the last line of (\ref{eq:ringxi2}) can be rewritten as
\begin{eqnarray} \label{eq:xistar1}
\mathring{\xi}_*^{\mu} &=& 2\kappa \mathring{\theta}^{\mu \nu} \mathring{B}_{\nu \rho \sigma} \mathring{\xi}_1^\rho \mathring{\xi}_2^\sigma +\kappa \mathring{\theta}^{\mu \sigma}{\cal \mathring{F}}_{\sigma \rho}^{\ \nu}\ (\mathring{\xi}_1^{\rho} \mathring{\lambda}_{2\nu}-\mathring{\xi}_2^{\rho} \mathring{\lambda}_{1\nu}) +\kappa^2 \mathring{\theta}^{\mu \nu} \mathring{\theta}^{\tau \sigma} {\cal \mathring{F}}^{\ \rho}_{ \nu \sigma}\ \mathring{\lambda}_{1\tau}\mathring{\lambda}_{2\rho} \\ \notag
&=& \kappa \mathring{\theta}^{\mu \nu} \mathring{\lambda}^*_{\nu}   \, .
\end{eqnarray}

Now relations (\ref{eq:Lstar1}) and (\ref{eq:xistar1}) define the $B\theta$-star bracket by
\begin{equation} \label{eq:star1}
[ \mathring{\Lambda}_1,\mathring{\Lambda}_2]^*_{\mathring{B} \mathring{\theta}} = \mathring{\Lambda}^* \Leftrightarrow [ (\mathring{\xi}_1, \mathring{\lambda}_1),(\mathring{\xi}_2, \mathring{\lambda}_2)]^*_{\mathring{B} \mathring{\theta}} = (\mathring{\xi}_*, \mathring{\lambda}^*) \, ,
\end{equation}
We can write the full bracket (\ref{eq:bracket4}) as
\begin{eqnarray} \label{eq:RoytZ}
[(\mathring{\xi}_1,\mathring{\lambda}_1),(\mathring{\xi}_2,\mathring{\lambda}_2)]_{{\cal C}_{\mathring{C}}} &=& \Big( [\mathring{\xi}_1,\mathring{\xi}_2]_L - [\mathring{\xi}_2,\mathring{\lambda}_1 \kappa \mathring{\theta}]_L + [\mathring{\xi}_1,\mathring{\lambda}_2 \kappa \mathring{\theta}]_L \\ \notag
&& + \frac{\kappa^2}{2} [\mathring{\theta},\mathring{\theta}]_S (\mathring{\lambda}_1,\mathring{\lambda}_2,.) , - [\mathring{\lambda}_1,\mathring{\lambda}_2]_{\kappa \mathring{\theta}} \Big) \\ \notag
&& + [(\mathring{\xi}_1,\mathring{\lambda}_1),(\mathring{\xi}_2,\mathring{\lambda}_2)]^*_{\mathring{B}, \mathring{\theta}} + [(\mathring{\xi}_1,\mathring{\lambda}_1),(\mathring{\xi}_2,\mathring{\lambda}_2)]^\star_{\mathring{\theta}} \, .
\end{eqnarray}

\subsection{Isotropic subspaces}

In order to give an interpretation to newly obtained starred brackets, it is convenient to consider isotropic subspaces. A subspace $L$ is isotropic if the inner product (\ref{eq:skalproizvod}) of any two generalized vectors from that sub-bundle is zero
\begin{equation} \label{eq:isodef}
\langle \Lambda_1, \Lambda_2 \rangle = 0, \ \ \ \Lambda_1 \, , \Lambda_2 \in L \, .
\end{equation}
From (\ref{eq:skalproizvod}), one easily finds that
\begin{eqnarray}\label{eq:isoth}
\xi_i^\mu = \kappa \, \theta^{\mu \nu} \lambda_{i \nu} . \qquad (i=1,2)   \qquad \theta^{\mu \nu} = - \theta^{\nu \mu} \, ,
\end{eqnarray}
for any bi-vector $\theta$, and
\begin{eqnarray}\label{eq:isoB}
\lambda_{i \mu}  = 2 B_{\mu \nu} \xi_i^\mu  . \qquad (i=1,2)   \qquad B_{\mu \nu} = - B_{\nu \mu} \, ,
\end{eqnarray}
for any 2-form $B$ satisfy the condition (\ref{eq:isodef}). 

Furthermore, it is straightforward to introduce projections on these isotropic subspaces by
\begin{eqnarray}\label{eq:Ithdef}
{\cal I}^\theta   (\Lambda^M) = {\cal I}^\theta  (\xi^\mu, \lambda_\mu) =  (\kappa \, \theta^{\mu \nu} \lambda_\nu  , \lambda_\mu ) \,  ,
\end{eqnarray}
and
\begin{eqnarray}\label{eq:IBdef}
{\cal I}_B   (\Lambda^M) = {\cal I}_B  (\xi^\mu, \lambda_\mu) =  (\xi^\mu, 2 B_{\mu \nu} \xi^\nu )  \,  .
\end{eqnarray}
Now it is easy to give an interpretation to star brackets. The $\theta$-star bracket (\ref{eq:star2}) can be defined as the projection of the Courant bracket (\ref{eq:CourantTheta}) on the isotropic subspace (\ref{eq:Ithdef})
\begin{equation} \label{eq:isostar2}
[\mathring{\Lambda}_1,\mathring{\Lambda}_2]^\star_{\mathring{\theta}} = {\cal I}^{\mathring{\theta}}\Big( [\mathring{\Lambda}_1, \mathring{\Lambda}_2]_{\cal C} \Big) \, .
\end{equation}

Similarly, note that all the terms in (\ref{eq:ringRLambda}) that do not appear in the $\theta$-twisted Courant bracket, contribute exactly to the $B\theta$-star bracket. From that, it is easy to obtain the definition of the $B\theta$-star bracket (\ref{eq:star1})
\begin{equation}
[\mathring{\Lambda}_1,\mathring{\Lambda}_2]^*_{\mathring{B}\mathring{\theta}} = {\cal I}^{\mathring{\theta}}\Big( [\mathring{\Lambda}_1, \mathring{\Lambda}_2]_{{\cal C}_{\mathring{C}}} \Big) - {\cal I}^{\mathring{\theta}}\Big( [\mathring{\Lambda}_1, \mathring{\Lambda}_2]_{{\cal C}_{\mathring{\theta}}} \Big) \, .
\end{equation}

\section{Courant bracket twisted by $B$ and $\theta$}
\cleq
Now it is possible to write down the expression for the Courant bracket twisted by $B$ and $\theta$ (\ref{eq:CTdef}), using the expression for $\mathring{C}$-twisted Courant bracket 
\begin{equation} \label{eq:CBCR}
[\breve{\Lambda}_1, \breve{\Lambda}_2]_{{\cal C}_{B\theta}} = A^{-1} [A \breve{\Lambda}_1,  A \breve{\Lambda}_2]_{{\cal C}_{\mathring{C}}}  \, ,
\end{equation}
where $A$ is defined in (\ref{eq:Adef}). Substituting (\ref{eq:CBCR}) into (\ref{eq:ringRxi}), we obtain
\begin{eqnarray} \label{eq:breveRxi}
\breve{\xi} &=& {\cal C}^{-1}[ {\cal C}\breve{\xi}_1,{\cal C}\breve{\xi}_2]_L - {\cal C}^{-1}[{\cal C}\breve{\xi}_2,\breve{\lambda}_1 \kappa {\cal C}^{-1}\mathring{\theta}]_L +{\cal C}^{-1}[{\cal C}\breve{\xi}_1,\breve{\lambda}_2 \kappa {\cal C}^{-1}\mathring{\theta}]_L \\ \notag
&&- \Big({\cal L}_{{\cal C}{\breve{\xi}_1}}(\breve{\lambda}_2{\cal C}^{-1}) - {\cal L}_{{\cal C}{\breve{\xi}_2}}(\breve{\lambda}_1{\cal C}^{-1}) - \frac{1}{2}d(i_{\breve{\xi}_1}\breve{\lambda}_2 - i_{\breve{\xi}_2}\breve{\lambda}_1)\Big)\kappa \mathring{\theta}{\cal C}^{-1}\\ \notag
&&+ \frac{\kappa^2}{2} {\cal C}^{-1} [\mathring{\theta},\mathring{\theta}]_S (\breve{\lambda}_1 {\cal C}^{-1},\breve{\lambda}_2 {\cal C}^{-1},.)  +\kappa {\cal C}^{-1} \mathring{\theta}\mathring{H}(., {\cal C}\breve{\xi}_1, {\cal C}\breve{\xi}_2) \\ \notag
&&-{\cal C}^{-1}\wedge^2 \kappa\mathring{\theta} \mathring{H}(\breve{\lambda}_1 {\cal C}^{-1}, ., {\cal C}\breve{\xi}_2) + {\cal C}^{-1}\wedge^2\kappa\mathring{\theta}\mathring{H}(\breve{\lambda}_2 {\cal C}^{-1}, ., {\cal C}\breve{\xi}_1)\\ \notag
&&+{\cal C}^{-1}\wedge^3\kappa \mathring{\theta}\mathring{H}(\breve{\lambda}_1 {\cal C}^{-1},\breve{\lambda}_2 {\cal C}^{-1},.) \, ,
\end{eqnarray}
and similarly, substituting (\ref{eq:CBCR}) into (\ref{eq:ringRLambda}), we obtain
\begin{eqnarray} \label{eq:breveRLambda}
\breve{\lambda} &=& \Big({\cal L}_{{\cal C}{\breve{\xi}_1}}(\breve{\lambda}_2{\cal C}^{-1}) - {\cal L}_{{\cal C}{\breve{\xi}_2}}(\breve{\lambda}_1{\cal C}^{-1}) - \frac{1}{2}d(i_{\breve{\xi}_1}\breve{\lambda}_2 - i_{\breve{\xi}_2}\breve{\lambda}_1)\Big){\cal C}  + \mathring{H}({\cal C}\breve{\xi}_1,{\cal C}\breve{\xi}_2,.){\cal C}\\ \notag
&&- [\breve{\lambda}_1 {\cal C}^{-1},\breve{\lambda}_2{\cal C}^{-1}]_{\kappa \mathring{\theta}}{\cal C}-\kappa\mathring{\theta} \mathring{H}(\breve{\lambda}_1{\cal C}^{-1}, ., {\cal C}\breve{\xi}_2){\cal C}+\kappa \mathring{\theta} \mathring{H}(\breve{\lambda}_2{\cal C}^{-1}, ., {\cal C}\breve{\xi}_1){\cal C} \\ \notag
&&+\wedge^2 \kappa\mathring{\theta} \mathring{H}(\breve{\lambda}_1{\cal C}^{-1}, \breve{\lambda}_2{\cal C}^{-1},.){\cal C} \, ,
\end{eqnarray}
where ${\cal C}^\mu_{\ \nu} = \Big( \cosh{\sqrt{\alpha}} \Big)^{\mu}_{\ \nu}$ and $\breve{\Lambda} = (\breve{\xi}, \breve{\lambda})$ (\ref{eq:Lxitilde}). This is somewhat a cumbersome expression, making it difficult to work with. To simplify it, with the accordance of our convention, we define the twisted Lie bracket by
\begin{equation} \label{eq:lietwist}
[\breve{\xi}_1, \breve{\xi}_2]_{L_{\cal C}} = {\cal C}^{-1}[ {\cal C}\breve{\xi}_1,{\cal C}\breve{\xi}_2]_L \, ,
\end{equation}
as well as the twisted Schouten-Nijenhuis bracket
\begin{equation} \label{eq:snbtwist}
\Big( [\breve{\theta}, \breve{\theta}]_{S_{\cal C}} \Big)^{\mu \nu \rho} = ({\cal C}^{-1})^{\mu}_{\ \sigma} ({\cal C}^{-1})^{\nu}_{\ \lambda} ({\cal C}^{-1})^{\rho}_{\ \tau} \Big( [{\cal C}\breve{\theta}, {\cal C} \breve{\theta}]_S \Big)^{\sigma \lambda \tau} \, ,
\end{equation}
and twisted Koszul bracket
\begin{equation} \label{eq:koszulwist}
[\breve{\lambda}_1, \breve{\lambda}_2]_{\theta_{\cal C}} =  ({\cal C}^T)^{-1} [ {\cal C}^T \breve{\lambda}_1 ,  {\cal C}^T \breve{\lambda}_2 ]_{\theta {\cal C}}  \, ,
\end{equation}
where the transpose of the matrix is necessary because the Koszul bracket acts on 1-forms. Now, the first three terms of (\ref{eq:breveRxi}) can be written as
\begin{equation} \label{eq:Lietr}
[ \breve{\xi}_1,\breve{\xi}_2]_{L_{\cal C}} - [\breve{\xi}_2,\breve{\lambda}_1 \kappa {\cal C}^{-1} \breve{\theta}]_{L_{\cal C}} +[\breve{\xi}_1,\breve{\lambda}_2 \kappa {\cal C}^{-1}\breve{\theta}]_{L_{\cal C}} \, ,
\end{equation}
where
\begin{equation} \label{eq:brevetheta}
\breve{\theta}^{\mu \nu} = ({\cal C}^{-1})^{\mu}_{\ \rho} \mathring{\theta}^{\rho \nu} = {\cal S}^{\mu}_{\ \rho} \theta^{\rho \nu} \, .
\end{equation}
The second line of (\ref{eq:breveRxi}) and the first line of (\ref{eq:breveRLambda}) originating from $\mathring{\theta}$ star bracket (\ref{eq:star2}) can be easily combined into
\begin{equation} \label{eq:star2tr}
[ ({\cal C} \breve{\xi}_1, \breve{\lambda}_1{\cal C}^{-1}),({\cal C} \breve{\xi}_2, \breve{\lambda}_2{\cal C}^{-1})]^\star_{{\cal C}^{-1}\breve{\theta}}\ {\cal C} \, .
\end{equation}
The terms originating from $\mathring{B} \mathring{\theta}$ star bracket (\ref{eq:star1}) are combined into
\begin{equation} \label{eq:star1tr}
 [(\breve{\xi}_1,\breve{\lambda}_1),(\breve{\xi}_2,\breve{\lambda}_2]^*_{\breve{B}, {\cal C}^{-1}\breve{\theta}} \, ,
\end{equation}
where
\begin{equation} \label{eq:breveB}
\breve{B}_{\mu \nu \rho} =  \mathring{B}_{\alpha \beta \gamma}{\cal C}^{\alpha}_{\ \mu} {\cal C}^{\beta}_{\ \nu} {\cal C}^{\gamma}_{\ \rho} = \Big( \partial_{\alpha} (B{\cal S}{\cal C}^{-1})_{\beta \gamma}+\partial_{\beta} (B{\cal S}{\cal C}^{-1})_{\gamma \alpha}+\partial_{\gamma} (B{\cal S}{\cal C}^{-1})_{\alpha \beta} \Big){\cal C}^{\alpha}_{\ \mu} {\cal C}^{\beta}_{\ \nu} {\cal C}^{\gamma}_{\ \rho} \, .
\end{equation}
The expressions for the Courant bracket twisted by both $B$ and $\theta$ can be written in a form
\begin{eqnarray}
[(\breve{\xi}_1, \breve{\lambda}_1),(\breve{\xi}_2, \breve{\lambda}_2)]_{{\cal C}_{B\theta}} &=& \Big( [ \breve{\xi}_1,\breve{\xi}_2]_{L_{\cal C}}  - [\breve{\xi}_2,\breve{\lambda}_1 \kappa {\cal C}^{-1} \breve{\theta}]_{L_{\cal C}} +[\breve{\xi}_1,\breve{\lambda}_2 \kappa {\cal C}^{-1}\breve{\theta}]_{L_{\cal C}}  \\ \notag 
&&+\frac{\kappa^2}{2}[\breve{\theta}, \breve{\theta}]_{S_{{\cal C}}}(\breve{\lambda}_1,\breve{\lambda}_2 ,.), -[\breve{\lambda}_1, \breve{\lambda}_2]_{\theta_{\cal C}} \Big) \\ \notag
&& +[ ({\cal C} \breve{\xi}_1, \breve{\lambda}_1{\cal C}^{-1}),({\cal C} \breve{\xi}_2, \breve{\lambda}_2{\cal C}^{-1})]^\star_{{\cal C}^{-1}\breve{\theta}}\ {\cal C} + [(\breve{\xi}_1,\breve{\lambda}_1),(\breve{\xi}_2,\breve{\lambda}_2]^*_{\breve{B}, {\cal C}^{-1}\breve{\theta}} \, .
\end{eqnarray}
When the Courant bracket is twisted by both $B$ and $\theta$, it results in a bracket similar to $\mathring{C}$-twisted Courant bracket, where Lie brackets, Schouten Nijenhuis bracket and Koszul bracket are all twisted as well.

\section{Conclusion}
\cleq

We examined various twists of the Courant bracket, that appear in the Poisson bracket algebra of symmetry generators written in a suitable basis, obtained acting on the double canonical variable (\ref{eq:Xdouble}) by the appropriate elements of $O(D,D)$ group. In this paper, we considered the transformations that twists the Courant bracket simultaneously by a 2-form $B$ and a bi-vector $\theta$. When these fields are mutually T-dual, the generator obtained by this transformation is invariant upon self T-duality.

We obtained the matrix elements of this transformation, that we denoted $e^{\breve{B}}$ (\ref{eq:ebb}), expressed in terms of the hyperbolic functions of a parameter $\alpha$ (\ref{eq:alfadef}). In order to avoid working with such a complicated expression, we considered another $O(D,D)$ transformation $A$ (\ref{eq:Adef}) and introduced a new generator, written in a basis of auxiliary currents $\mathring{\iota}_\mu$ and $\mathring{k}^\mu$. The Poisson bracket algebra of a new generator was obtained and it gave rise to the $\mathring{C}$-twisted Courant bracket, which contains all of the fluxes. 

The generalized fluxes were obtained using different methods \cite{royt, nick1, nick2,flux1, flux2, flux3}. In our approach, we started by an $O(D,D)$ transformation that twists the Courant bracket simultaneously by a 2-form $B$ and bi-vector $\theta$, making it manifestly self T-dual. We obtained the expressions for all fluxes, written in terms of the effective fields 
\begin{equation} \label{eq:Bmathring1}
\mathring{B}_{\mu \nu} = B_{\mu \rho}  \Big( \frac{\tanh{\sqrt{2 \kappa \theta B}}}{\sqrt{2 \kappa \theta B}} \Big)^\rho{}_\nu \, , \qquad
 \mathring{\theta}^{\mu \nu} = \Big( \frac{\sinh{2  \sqrt{2 \kappa \theta B}}}{2 \sqrt{2 \kappa \theta B}} \Big)^\mu{}_\sigma \theta^{\sigma \nu}  \, .
\end{equation}
The fluxes, as a function of these effective fields, appear naturally in the Poisson bracket algebra of such generators. 

Similar bracket was obtained in the algebra of generalized currents in \cite{nick1, nick2} and is sometimes referred to as the Roytenberg bracket \cite{royt}. In that approach, phase space has been changed, so that the momentum algebra gives rise to the $H$-flux, after which the generalized currents were defined in terms of the open string fields. The bracket  obtained this way corresponds to the Courant bracket that was firstly twisted by $B$ field, and then by a bi-vector $\theta$. The matrix of that twist is given by
\begin{equation} \label{eq:enaR}
e^R = e^{\hat{\theta}} e^{\hat{B}} = \begin{pmatrix}
\delta^\mu_\nu + \alpha^\mu_{\ \nu} & \kappa \theta^{\mu \nu} \\
2 B_{\mu \nu} & \delta^\nu_\mu
\end{pmatrix}\, .
\end{equation}
In our approach, we obtained the transformations that twists the Courant bracket at the same time by $B$ and $\theta$, resulting in a $\mathring{C}$-twisted Courant bracket. As a consequence, the $\mathring{C}$-twisted Courant bracket is defined in terms of auxiliary fields $\mathring{B}$  (\ref{eq:Bmathring}) and $\mathring{\theta}$ (\ref{eq:thetamathring}), that are themselves function of $\alpha$. This is not the case in \cite{nick1, nick2}. The Roytenberg bracket calculated therein can be also obtained following our approach by twisting with the matrix
\begin{equation} \label{eq:eR}
e^C =A e^{\breve{B}} = 
\begin{pmatrix}
{\cal C}^2 & \kappa ({\cal CS}\theta) \\
2B{\cal CS} & 1 
\end{pmatrix}
\, ,
\end{equation}
demanding that the background fields are infinitesimal $B \sim \epsilon$, $\theta \sim \epsilon$ and keeping the terms up to $\epsilon^2$. With these conditions, $e^C$ (\ref{eq:eR}) becomes exactly $e^R$ (\ref{eq:enaR}), and the bracket becomes the Roytenberg bracket.

Analyzing the $\mathring{C}$-twisted Courant bracket, we recognized that certain terms can be seen as new brackets on the space of generalized vectors, that we named star brackets. We demonstrated that they are closely related to projections on isotropic spaces. It is well established that the Courant bracket does not satisfy the Jacobi identity in general case. The sub-bundles on which the Jacobi identity is satisfied are known as Dirac structures, which as a necessary condition need to be subsets of isotropic spaces. Therefore, the star brackets might provide future insights into integrability conditions for the $\mathring{C}$-twisted Courant bracket \cite{sorad}.

In the end, we obtained the Courant bracket twisted at the same time by $B$ and $\theta$ by considering the generator in the basis spanned by $\breve{\iota}$ and $\breve{k}$, equivalent to undoing $A$ transformation, used to simplify calculations. With the introduction of new fields $\breve{B}_{\mu \nu}$ and $\breve{\theta}^{\mu \nu}$, this bracket has a similar form as $\mathring{C}$-twisted Courant bracket, whereby the Lie, Schouten-Nijenhuis and Koszul brackets became their twisted counterparts. 

It has already been established that $B$-twisted and $\theta$-twisted Courant brackets appear in the generator algebra defined in bases related by self T-duality \cite{crdual}. When the Courant bracket is twisted by both $B$ and $\theta$, it is self T-dual, and as such, represent the self T-dual extension of the Lie bracket that includes all fluxes. It has been already shown \cite{cdual} how the Hamiltonian can be obtained acting with $B$-transformations on diagonal generalized metric. The same method could be replicated with the twisting matrix $e^{\breve{B}}$, that would give rise to a different Hamiltonian, whose further analysis can provide interesting insights in the role that the Courant bracket twisted by both $B$ and $\theta$ plays in understanding T-duality.

\end{document}